\documentclass[12pt]{iopart}

\usepackage{float}
\usepackage{amsfonts}
\usepackage{amssymb}
\usepackage{framed}
\usepackage{graphicx}
\usepackage{caption}
\usepackage{subcaption}
\usepackage{iopams}
\usepackage{setstack}
\usepackage{cite}

\newcommand\beq{\begin{equation}}      
\newcommand\beqnn{\begin{eqnarray*}}   
\newcommand\beqa{\begin{eqnarray}}     
\newcommand\beqann{\begin{eqnarray*}}  

\newcommand\eeq{\end{equation}}        
\newcommand\eeqnn{\end{eqnarray*}}     
\newcommand\eeqa{\end{eqnarray}}       
\newcommand\eeqann{\end{eqnarray*}}    

\def\nl {\nonumber \\}

\def\tpm {T_{+-}} \def\tmp {T_{-+}} \def\tpp {T_{++}} \def\tmm {T_{--}}
   
\def\tc {T_{c}} \def\ts {T_{s}}

\begin{document}

\title{Conservation of polymer winding states: a combinatoric approach}
\author{C.M. Rohwer, K.K. M\"uller-Nedebock, F.-E. Mpiana Mulamba}
\address{Institute of Theoretical Physics, University of Stellenbosch, Stellenbosch 7600, South Africa}
\ead{crohwer@sun.ac.za}

\begin{abstract}
The work in this article is inspired by a classical problem: the statistical physical
properties of a closed polymer loop that is wound around a rod.  Historically the preserved
topology of this system has been addressed through identification of similarities with 
magnetic systems.  We treat the topological invariance in terms of a set of rules that
describe all augmentations by additional arcs of some fundamental basic loop of a given winding
number.  These augmentations satisfy the Reidemeister move relevant for the polymer with 
respect to the rod.  The topologically constrained polymer partition function is now constructed 
using the combinatorics of allowed arc additions and their appropriate statistical weights.
We illustrate how, for winding number 1, we can formally derive expressions for lower and
upper bounds on the partition function.  Using the lower bound approximation we investigate 
a flexible polymer loop wound between two slits, calculating the force on the slit as well as
the average numbers of arc types in dependence of slit width and separation.  Results may be extended to higher winding numbers.  
The intuitive nature of this combinatoric scheme allows the development of a variety of approximations and
generalisations.

\end{abstract}
\pacs{61.41.+e, 36.20.Ey, 36.20.Fz}

\maketitle

\section{Introduction}

Entanglements occur naturally in polymeric systems, and are related to topological constraints. Indeed, the fact that different polymer strands cannot pass through each other is manifest in two observations: a) there exists an excluded volume interaction in real polymers that leads to self-avoidance, and b) for closed loops topological states must be conserved. We will focus on the latter statement, and concern ourselves with the associated configurational constraints. Such constraints determine which configurations of the polymeric system are topologically equivalent to each other, and thus restrict the polymer configurations over which we must sum to calculate the partition function for a given topological state. Mathematically the notion of topological equivalence may be captured (at least partially) by topological invariants. The role of topological constraints in polymers remains an important issue in various systems 
(see reviews by Kholodenko and Vilgis\cite{kholodenko1998}, and, more recently,
\cite{Orlandini2007,Grosberg2009,Micheletti2011}).  Entanglement of synthetic 
or biological macromolecule loops continues to be treated in analytical and computational modelling (for example, in \cite{Marko2009,Marko2010,Bohn2010}) and is deemed to be particularly relevant for localisation of DNA 
in cells \cite{Marenduzzo2010}.  Although computer simulations have driven results strongly, 
there is still a need for expanding the range of
analytical tools to deal with entangled chains.

As early as 1961, Frisch and Wasserman \cite{frischwasserman1961} considered topological isomerism in chemical systems, investigating knotted and unknotted links, loops and rings that occur in certain chemical molecules. Soon thereafter (1967), Edwards \cite{edwards1967} explicitly pointed out the importance of topological constraints to polymeric systems, and that such constraints need to be included into the statistical mechanics of polymers. In that article, the specific example of a polymer wound around a rod is considered, and the winding number is identified as a suitable topological invariant that categorises topologically distinct configurations. The winding number simply represents the number of times the polymer winds around the rod before closing on itself. Viewing this problem in a projection along the rod, one may capture the planar winding number (angle) of the polymer around a fixed point (i.e., the projection of the rod) through the integral
\beq
\oint  \frac{x \dot{y} - y \dot x}{x^2+y^2}\,ds,
\eeq
where $s$ is the arc-length parameter of the strand, and $\dot x \equiv \frac{\partial x}{\partial s}$ etc. Naturally this winding number cannot be altered once the polymer loop has been wound around the rod and closed. In \cite{edwards1967} this constraint is included into the path integral partition function (probability distribution) of the polymer inside a delta function that is exponentiated through the introduction of auxiliary fields. The resulting action is related to that of a magnetic system, and is treated under certain approximations. In the same year as Edwards, the same physical system was investigated by Prager and Frisch \cite{pragerfrisch1967}. This problem can also be addressed through the introduction of a tailored potential that approximates an interaction at short distances with the rod, and is included in the path integral through an additional Boltzmann factor \cite{wiegel1983,wiegel1986}. We shall, however, focus on topological invariance as the basis of our approach.

A further simple invariant is the linking number, which tells us how many times two distinct loops $a$ and $b$ are wound around each other. The linking number is given by the Gauss integral
\beq
\oint_a \oint_b \left( \dot {\vec r}_a(s_a) \times \dot {\vec r}_b(s_b)\right) \cdot \frac{\vec r_a - \vec r_b}{|\vec r_a - \vec r_b|^3}\;ds_a \,ds_b,
\eeq
where the position vectors $\vec r$ and their derivatives $\dot{\vec r}$ are parametrised by the arc-lengths $s_a$ and $s_b$. This invariant may also be included as a constraint in the polymer path integral (see, for instance, \cite{edwards1967,breretonshah1980,kholodenko1998,brereton1995,breretonvilgis1995}). 

Winding and linking numbers (and other basic invariants) do not capture all topological constraints in polymer systems. More complex higher order invariants (which may represent more detailed topological information) exist and may, in principle, also be included as constraints in path integrals. Indeed, the field theories that thus arise from topologically constrained polymer systems have been studied extensively \cite{kholodenko1998,breretonshah1980,brereton1995,breretonvilgis1995}. There is also a deep connection between polynomial knot invariants and quantum field theories. The reader is referred to the seminal work by Witten \cite{witten1989} where Jones polynomials are investigated in the setting of Yang-Mills theory. This work opened the door to subsequent extensions such as perturbative approaches -- see, for instance, \cite{vanderwetering1992}. 

Typically, however, simple invariants such as winding numbers and linking numbers have been considered in the context of polymer path integrals since more complex invariants become mathematically tedious to handle. The conservation of these topological invariants has been shown to relate to symmetry transformations that ultimately manifest in local gauge invariance in such field theories \cite{breretonshah1980,brereton1995}.

In this article we shall address the often studied problem of a polymer wound around an infinitely long obstacle in a plane. In doing so, we shall not consider any self-entanglements of the loop, but simply concern ourselves with the topology of the loop \emph{relative to the rod}. Consequently the mathematical intricacies of higher order invariants will not be of bearing here: we need look no further than winding numbers to address this physical system. Perhaps it is (in part) for this reason that the ``polymer wound around a rod'' has been studied so extensively. \textbf{More recently (in 2003) Grosberg and Frisch \cite{grosbergfrish2003} presented various modifications and extensions of Edwards' original results, both in the quenched (constrained partition sum) and the annealed (probability distributions of winding angles) settings. These include confining the polymer-and-rod system to a cavity, and winding the polymer around a disc.} Our aim in this article is similar that of \cite{edwards1967} and \cite{pragerfrisch1967}: we wish to find the partition function of a polymer wound around an obstacle. \textbf{Our partition function, obtained through a different calculational approach, is then used to study physical quantities for various geometries.}

\textbf{We employ a strategy that differs significantly from the path integral schemes cited above, namely to evaluate invariant knots by a combinatoric scheme. This may be done in terms of enumerations of braids, as shown by Nechaev and co-workers \cite{nechaev1996,nechaev1999}.  We also develop a combinatoric scheme, and use this for enumerating configurations subject to a winding number constraint in particular. 
In principle the sequence of allowed topological configurations may be generated braid-theoretically.  Since we, however, wish to calculate the partition function of polymer degrees of freedom and related quantities, it further is necessary to couple these degrees of freedom in a more careful manner to the combinatorics.  This is presented in two parts.  Firstly we outline how configurations may be labelled according to piercings that the strand makes through a plane.  Then we consistently approximate the sum of all unique but topologically equivalent configurations comprised of connected arcs restricted to half-space.  This method avoids the complications brought about by delta-function constraints which would be necessary to enforce the winding number in various alternate scenarios.
This enables us to find upper and lower bounds for the free energy. In principle, this formalism allows for the description of various types of polymer chains, e.g., Gaussian or semi-flexible. 
} Secondly, we illustrate the calculation of the partition function and related averages for various winding scenarios, e.g., winding between two slits. We are able to calculate forces and ratios of arc types for confining geometries. To this end we consider Gaussian chains and their associated probability distributions, and calculate statistical quantities of interest.

\section{Winding a polymer around a rod}
\label{windingsection}
The basis of our problem is a topological obstacle around which a polymer strand is wound. We start by presenting some simple examples of loops wound around a rod, 
\textbf{and then illustrate different configurational modifications / augmentations that do not alter the topology of these basic loops. Such procedures are essentially braid manipulations on two strands which may be represented in terms of braid groups -- see, for instance \cite{kauffman,manturov}. Since we immediately couple the configurations of this quenched scenario to polymer degrees of freedom, however, the braid group relations alone are not sufficient for our enumeration procedure. This is discussed in detail in \ref{braidappend}, in reference to \cite{nechaev1996,nechaev1999}.}

\subsection{Example of the basic loop, winding number $w=1$}
\label{basicsection}
\begin{figure}[H]
	\centering
		\includegraphics[width=0.60\textwidth]{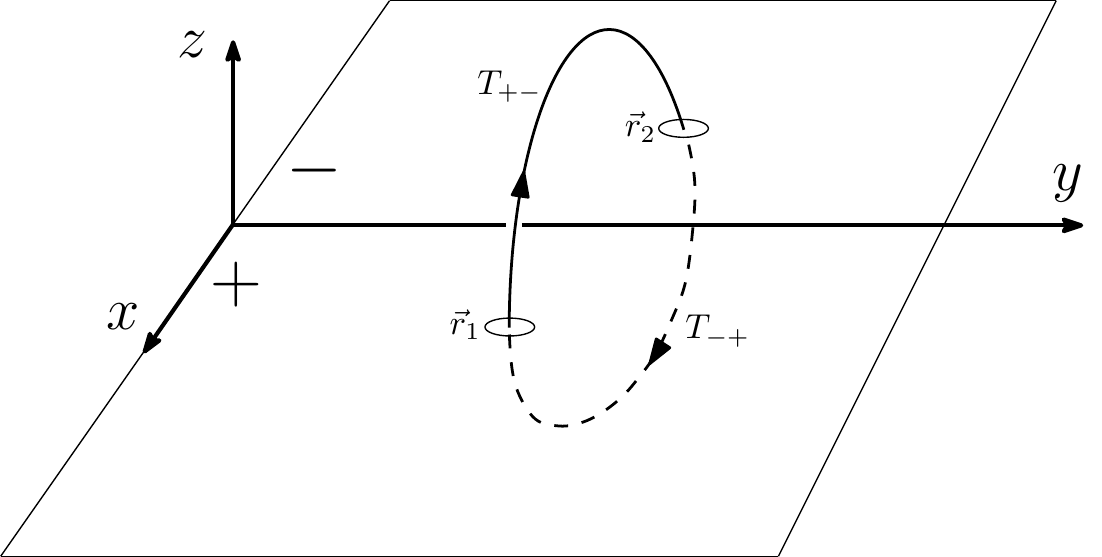}
	\caption{Closed polymer loop, $w$ = 1.}
	\label{fig:fig1}
\end{figure}

Consider an infinitely long rod that is placed along the $y$ axis of a system of axes in $\mathbb R ^3$. Suppose now that an open polymer strand is wound around this rod and then closed on itself to form a closed loop. The first natural question to ask is \textit{how often} the polymer is wound around the rod. Indeed, this number distinguishes topologically distinct configurations of the polymer, and is appropriately known as the winding number, $w$.  For the remainder of the article, the scenario in \Fref{fig:fig1} with the minimal number of arcs will be referred to as the \emph{basic loop}. If we take this configuration with $w=1$ (the simplest case where strand is wrapped around the rod only once) we cannot deform or alter this configuration to obtain one where $w\neq1$ without physically breaking the polymer strand. In this sense the winding number is a topological invariant of the particular configuration created when closing the open strand after $w$ windings.
Let us divide the complete polymer loop into sub-arcs, the division occurring whenever the $xy$-plane is pierced by the polymer. For the example in Figure \ref{fig:fig1} the entire polymer loop may be viewed as consisting of a ``sequence'' of two polymer arcs, each constrained by the $xy$-plane to a half-space w.r.t. the positive / negative $z$ axis. In this figure we have labelled each of these arc-segments with a $T$ whose subscript is $+-$ if the arc begins in the $x>0$ half-plane $\{(x,y):\, x>0,\,y\in\mathbb R\}$ and ends in the $x<0$ half-plane $\{(x,y):\, x<0,\,y\in\mathbb R\}$. The subscript $-+$ applies to the opposite case. The $T$s themselves will later represent the probability distributions for the half-space restricted sub-arcs. They are functions of the (planar) beginning and end co-ordinates of the respective polymer segments and of the arc-lengths of the segments. For now we will simply use this notation to represent sequences of such sub-arcs, and demonstrate how we may capture the topology of any polymer loop as a composition / sequence of such $T$s. In this spirit, we represent the simple closed loop of Figure \ref{fig:fig1} symbolically by the sequence $\tpm \tmp$ (or alternatively by the cyclic permutation $\tmp \tpm$). Essentially the sequence of subscripts indicates how one would follow the polymer strand around the rod from one piercing of the plane to another. An orientation convention (indicated by the arrows in \Fref{fig:fig1}) is chosen without loss of generality. The partition function corresponding to \Fref{fig:fig1} in less compact notation would be
\beq
Z = \int_{\mathcal{D}} \tpm(\vec r_1,\vec r_2)\,\tmp(\vec r_2,\vec r_1) = \int_{\mathcal{D}} \tmp(\vec r_2,\vec r_1)\,\tpm(\vec r_1,\vec r_2),
\label{zfig1}
\eeq
where the position vectors label the piercings of the $xy$-plane and integration is over the relevant domain
\beq
\mathcal D = \{x_1 \in (0,\infty);\; x_2 \in (0,-\infty);\;y_1,y_2\in(-\infty,\infty)\}.
\eeq
Each $T(\vec r_i, \vec r_j)$ represents the statistical weight of a polymer arc restricted to half-space with appropriate initial and final positions in the plane. $T$ depends on the nature of the specific polymer. In principle the methods shown here can be applied to Gaussian as well as semiflexible polymers, etc. The symbolic sequence $\tpm \tmp$ thus represents the integrand of the partition function for the basic loop. This notion will be clarified in \sref{partitionsection}. We shall now extend this picture (and the symbolic notation) to higher winding numbers.

\subsection{Higher winding numbers: $w>1$}
\label{w2section}
\begin{figure}[H]
	\centering
		\includegraphics[width=0.60\textwidth]{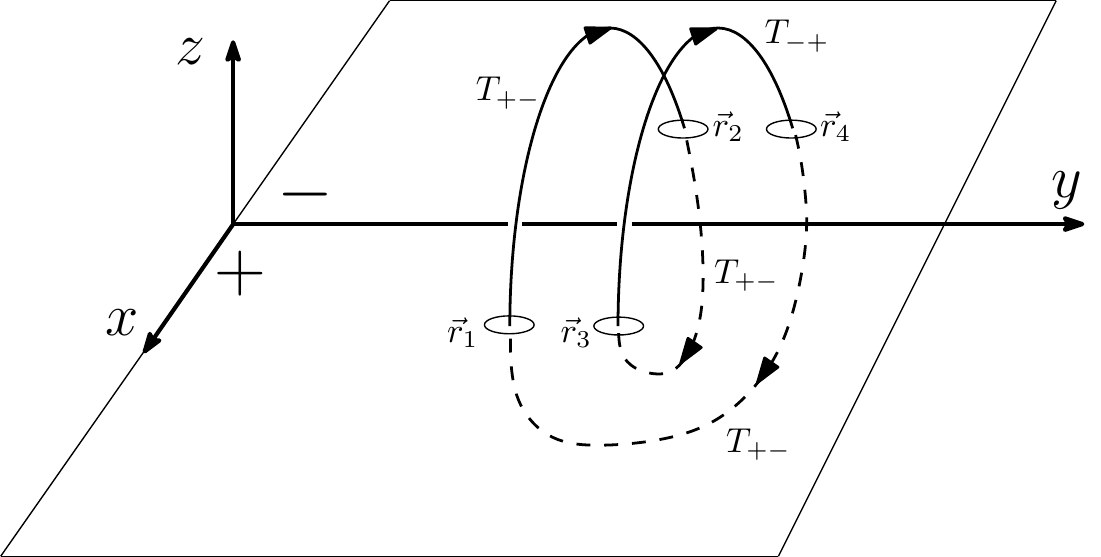}
	\caption{Closed polymer loop, $w$ = 2.}
	\label{fig:w2}
\end{figure}

Figure \ref{fig:w2} shows a polymer strand with winding number $w=2$.
Again the complete polymer strand may be decomposed into a sequence of confined sub-arcs, each living in either the $z>0$ or the $z<0$ half-space. Following the notation above, the case $w=2$ can thus be described symbolically by the sequence $\tpm \tmp \tpm \tmp$ (or any cyclic permutation thereof, depending on the choice of reference point for labeling). The partition function for Figure \ref{fig:w2} would be
\beqa
Z &=& \int_{\mathcal D} \tpm(\vec r_1,\vec r_2)\,\tmp(\vec r_2,\vec r_3) \,\tpm(\vec r_3,\vec r_4)\,\tmp(\vec r_4,\vec r_1) \nl
&=& \int_{\mathcal D} \tmp(\vec r_2,\vec r_3) \,\tpm(\vec r_3,\vec r_4)\,\tmp(\vec r_4,\vec r_1)\, \tpm(\vec r_1,\vec r_2)\quad\mathrm{etc.}
\label{zfig2}
\eeqa
Clearly Z is invariant under cyclic permutation of the factors in the integrand in equation \eref{zfig2}. Consequently one could just as well label \Fref{fig:w2} with any cyclic permutation of $\tpm \tmp \tpm \tmp$. The order of the $T$s does, however, matter, since the arguments of consecutive $T$s (i.e., $\vec r_i$, $\vec r_{i+1}$ etc.) must match up. This may be viewed in analogy to operator multiplication. 

The strategy in \sref{partitionsection} will be to ``diagonalise'' the integral above, so that we may write symbolically $\tpm \tmp \tpm \tmp = (\tpm \tmp)^2$. Analogously loops wound $w$ times are expressed as $(\tpm \tmp)^w$ in this compact notation. 

The examples considered thus far only show limited configurations associated with specific winding numbers, since they are composed of sub-arcs that cross from one side of the rod to the other. Other permissible configurations (that maintain the winding number) can include sub-arcs that remain on one side of the rod. In the next section we illustrate how simple loops such as those in Figures \ref{fig:fig1} and \ref{fig:w2} may be augmented in this way. Some combinatoric rules will be established on the symbolic level of sequences of $T$s. The connection of these combinatoric sequences to a complete partition function will be made in \sref{partitionsection}. We shall now focus on the case of $w=1$, since the partition functions for higher winding numbers are generated from powers of the basic loop.

\subsection{Augmenting the basic loop: insertion of sub-arcs}

\begin{figure}[H]
	\centering
		\includegraphics[width=0.6\textwidth]{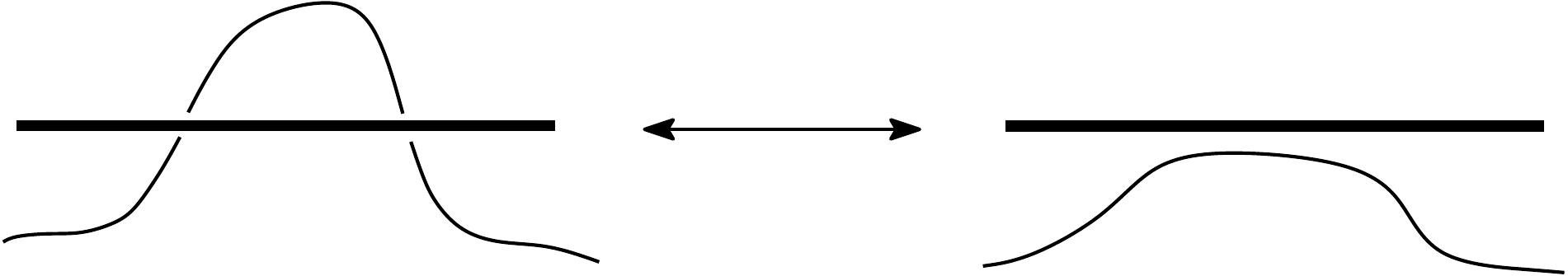}
	\caption{Type two Reidemeister move of the polymer (thin) relative to the rod (thick).}
	\label{fig:moveR2}
\end{figure}

Let us return to the basic loop from Figure \ref{fig:fig1} with $w$ = 1. We note that it is possible to augment or ``decorate'' this simple loop with more half-space constrained sub-arc segments. This process is subject to a Reidemeister move of the second type (see Figure \ref{fig:moveR2}) of the polymer strand relative to the rod, viewed in a side-on projection along the $x$ axis. 
Since the Reidemeister moves do not alter a particular topological state for closed strands \cite{reidemeister}, this augmentation does not alter the winding number but simply introduces more piercings of the $xy$-plane. Clearly the number of $T$s equals the number of piercings for a particular configuration. Introducing additional piercings / arc-segments may be done in two ways:

\begin{enumerate}
	\item We may insert two more sub-arcs, each beginning and ending in the half-plane $\{(x,y):\, x>0,\,y\in\mathbb R\}$, one living in half-space $z<0$ and the other in $z>0$:
\begin{figure}[H]
	\centering
		\includegraphics[width=0.60\textwidth]{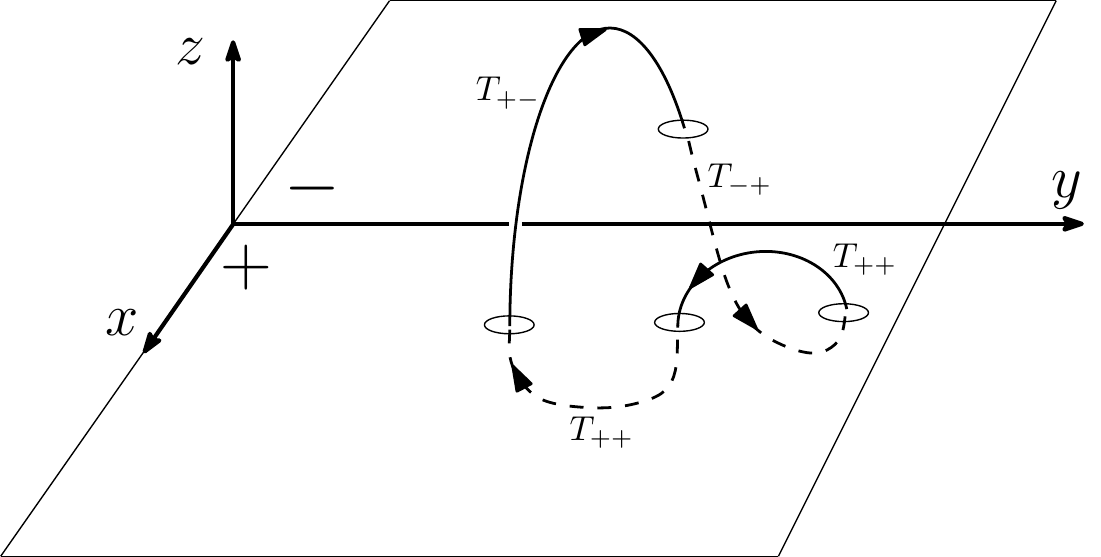}
	\caption{Closed polymer loop, $w$ = 1, additional constrained arc-segments.}
	\label{fig:fig2}
\end{figure}
We describe the sequence of arcs in Figure \ref{fig:fig2} with the symbolic sequence $\tpm \tmp \tpp \tpp$ (or any cyclic permutation). Here the subscript $++$ indicates a sub-arc beginning and ending in the half-plane $\{(x,y):\, x>0,\,y\in\mathbb R\}$.

\item Given Figure \ref{fig:fig2} we observe that we could also take one of the $\tpp$ sub-strands and ``pull it across'' the rod, as follows:
\begin{figure}[H]
	\centering
		\includegraphics[width=0.60\textwidth]{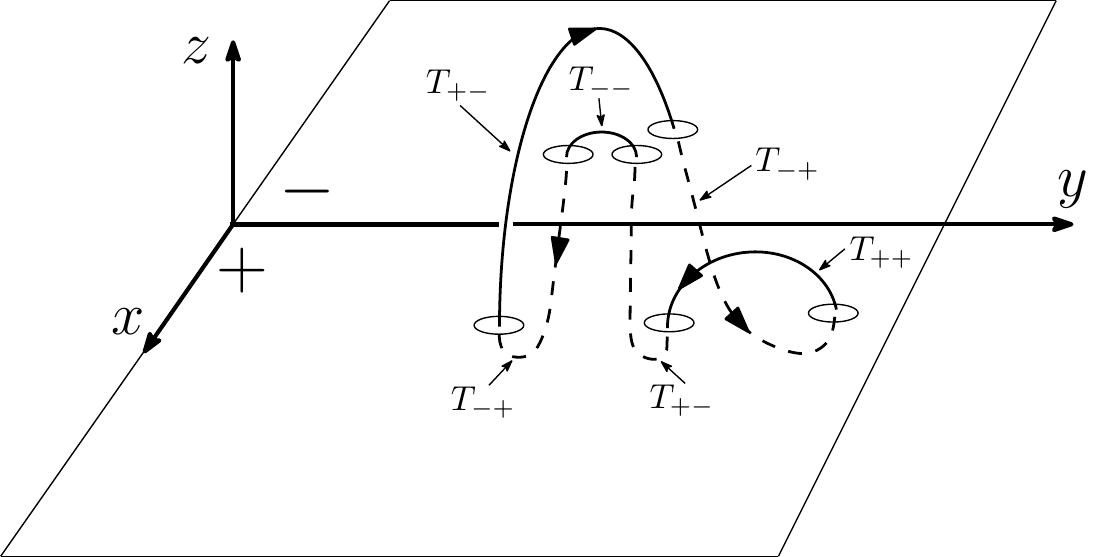}
	\caption{Closed polymer loop, $w$ = 1, a further augmentation.}
	\label{fig:fig3}
\end{figure}
The diagram shown here could be described by the sequence $\tpm \tmp \tpp \tpm \tmm \tmp$ (or any cyclic permutation thereof).

\end{enumerate}
These two augmentation procedures form the basis for a set of combinatoric rules that govern what sequences are derivable from the basic loop with $w=1$. Let us return once more to said basic loop in \Fref{fig:fig1}. From the examples above it is clear that this is the simplest loop for two reasons: (i) it has the smallest possible winding number $w=1$, and (ii) it has the smallest number of piercings of the $xy$-plane. Clearly the two examples in Figures \ref{fig:fig2} and \ref{fig:fig3} are topologically equivalent to that in \Fref{fig:fig1}: they have the same winding number. However, they have more piercings of the $xy$-plane because more sub-arcs were inserted. The inclusion of these augmenting sub-arcs was topologically consistent: the winding number was not altered and the strand was not broken. We continue now by stating concretely what rules govern the augmentation of basic loops through insertion of sub-arcs in such a way that the basic topology (i.e., their winding numbers) is conserved.

\subsection{Condensed notation}
\label{condensednotation}
The statistical weights of the sub-arcs considered here are symmetric around $z=0$. It is, however, important to distinguish between $T$s that cross the rod and those whose two piercings of the plane are on the same side of the rod. We introduce the following shorthand,
\beq
\tpm, \tmp \rightarrow T_c
\eeq
and
\beq
\tpp, \tmm \rightarrow T_s,
\eeq
where the subscripts $c$ and $s$ refer to ``crossing'' and ``same side'', respectively. For Figure \ref{fig:fig3}, for example, we may write
\beq
\tpm \tmp \tpp \tpm \tmm \tmp \;=\; T_c T_c T_s T_c T_s T_c.
\label{egstring} 
\eeq
Of course it is implicit that a string be uninterrupted in its subscripts: two consecutive $T$s of the form
\beq
T_{\alpha \beta} T_{\gamma \delta}\quad \mathrm{with} \quad\alpha, \beta,\gamma, \delta\in\{+,-\}
\eeq
must be such that $\beta=\gamma$. If this were not the case one would have a broken strand since consecutive sub-arcs in different half-planes cannot be connected due to the rod which separates the two half-planes.

\subsection{Augmentation rules: maintaining $w=1$}
\label{rules}
It should be noted that the condition of continuity in subscripts between consecutive $T$s (as set out in section \ref{condensednotation}) is not sufficient to ensure that any polymer loop described by a string of $T$s with this unbroken property need be topologically equivalent to the basic loop with $w=1$. This is easily seen from the string $(T_c T_c)^w$, $w>1$, which is clearly unbroken, but is not topologically equivalent to the case $w=1$. Indeed, only strings that are derived from each other in very specific ways represent the same topology. To illustrate this, we now summarise some elementary inferences derived from the examples above:

\begin{enumerate}
	\item A closed loop with $w=1$ in its simplest form (i.e., with the minimal number of piercings of the $z=0$ plane) is represented by the sequence 
	\beq
	\tc \tc.
	\eeq
	\item A closed loop with $w>1$ in its simplest form (i.e., with the minimal number of piercings of the $z=0$ plane) is represented by the sequence 
	\beq
	(\tc \tc)^w.
	\label{biggerw}
	\eeq
	\item The basic sequence $\tc\tc$ can be augmented (``dressed'') as in Figure \ref{fig:fig2} according to the lengthening rule
	\beq
	\tc \longrightarrow \tc \ts \ts.
	\label{rulea}
	\eeq
	\item A further augmentation procedure, as shown in Figure \ref{fig:fig3} is described by the replacement rule
	\beq
	\ts \longrightarrow \tc \ts \tc.
	\label{ruleb}
	\eeq
\end{enumerate}

It is clear that compound rules arise, namely
	\beq
	T_x \longrightarrow T_x \ts \ts
	\label{ruleanew}
	\eeq
and
	\beq
	\ts \longrightarrow (\tc)^n \ts (\tc)^n.
	\label{rulebnew}
	\eeq
\textbf{We note here that (\ref{ruleanew}) and (\ref{rulebnew}) essentially encode group relations of the braid group $B_2$ -- see \ref{braidappend}.}
Any other alteration of the polymer strand through insertion / alteration of $T$s that is \textit{not} of type \eref{rulea} or \eref{ruleb} (or equivalently \eref{ruleanew} or \eref{rulebnew}) would necessarily either break the strand (see previous section) or increase the winding number (see \eref{biggerw}). This implies that the two rules \eref{ruleanew} and \eref{rulebnew} above capture all possible ways of generating loops that are topologically equivalent to the basic loop shown in Figure \ref{fig:fig1}. As in sections \ref{basicsection} and \ref{w2section}, calculation of the partition function requires integration over various degrees of freedom.

\subsection{What sequences are valid for $w=1$?} 
\label{validseq}
Valid sequences generated from the simplest form $\tc\tc$ ($w = 1$) according to \sref{rules} must have the following properties:
	\begin{enumerate}
	\item in order for the loop to be closed, the first and last index must be equal (where cyclic permutations of sequences are equivalent) - see equation (\ref{egstring}) as an example,
	\item for the same reason, the second index of any $T$ in the sequence must equal the first index of the next $T$,
	\item the total number of $T$s in the sequence must be even (since the basic undecorated closed loop has two terms, and both augmentation rules keep the total number even),
	\item the number of $\tc$s must be even,
	\item the number of $\ts$s must be even,
	\item the string must be algorithmically reducible (this is defined in the next section).
\end{enumerate}

\subsection{Algorithmic reducibility of valid strings for $w=1$}
\label{reducibilitysect}
Let us define the sets of generic functional units / substrings counting either even ($G$) or odd ($U$) sequences of $T_s$:
\beq
G_n = \tc (\ts)^{2n},\;\;n=0,1,2,\ldots
\eeq
and
\beq
U_n = \tc (\ts)^{2n+1},\;\;n=0,1,2,\ldots.
\eeq
Any string of $T$s could now be rewritten as a string of $G$s and $U$s. For the basic unit for $w=1$ we may write $T_c T_c = G_0 G_0$, with another example being $T_c T_c T_s T_s T_s T_c T_s T_c T_s T_s = G_0 U_1 U_0 G_1$. Since the substring $T_s T_s$ may be trivially inserted or removed in any sequence (see \eref{ruleanew}), it is clear that
\beq
G_n \leftrightarrow G_0
\label{gnrule}
\eeq
and
\beq
U_n \leftrightarrow U_0.
\label{unrule}
\eeq
We further infer from \eref{ruleanew} and \eref{rulebnew} that
\beq
X\,U_0^{2n}\,Y \leftrightarrow XY  \quad (n = 1,2,\ldots) \quad \forall\;\;\mathrm{substrings}\; \;X,Y
\label{functionalrulea}
\eeq
and
\beq
G_0^{m}\, U_0 \,G_0 ^m \leftrightarrow U_0 \quad (m = 1,2,\ldots).
\label{functionalruleb}
\eeq
We define a given string to be algorithmically reducible if the following procedure is possible:
\begin{enumerate}
	\item apply \eref{gnrule} and \eref{unrule} to simplify the string wherever possible,
	\item apply  \eref{functionalrulea} to simplify the string wherever possible,
	\item now apply \eref{functionalruleb} to simplify the string wherever possible,
	\item repeat until only the functional unit $G_0 G_0$ remains.
\end{enumerate}
Strings that are algorithmically reducible in this manner are topologically equivalent to the basic unit $\tc \tc$ which represents $w=1$. For $w>1$ the string $(G_0)^{2w}$ would remain in step (iv) after complete application of this procedure.

\section{Partition function}
\label{partitionsection}

The full partition function for a given winding number is now given by the integrals over the sums of all the configurations that are compatible with the winding number.

\subsection{Summing over diagrams}
\label{sumsection}
The rules by which moves are produced do lead to all possible configurations permissible as described in \sref{windingsection}. The corresponding sequence of $T$s represents the statistical weight for each configuration. In order to enumerate the valid sequences correctly, each distinct configuration needs to occur exactly once in the partition function.  (Alternatively one needs to be able to determine the correct multiplicity for the crossings in order to sum the appropriate terms in the partition function correctly.)  

For completeness we state here once more the rules from equations \eref{ruleanew} and \eref{rulebnew},
\begin{itemize}
\item[(i)] $T_x \rightarrow  T_x T_s T_s$ and
\item[(ii)] $T_s \rightarrow T_c T_s T_c$.
\end{itemize}
The first rule adds loops of the type $T_s$ in \emph{even multiples} and the second rule is responsible for the addition of new terms in $T_c$.  It is simple to see that different sequences of applying the rules (i) and (ii) above, on different elements, can lead to configurations that are identical. This has obvious implications in writing expressions for the sum in the partition function.  Here we investigate whether a scheme by which enumeration or an approximate enumeration are possible.  (The explicit procedure can be compared to the configurations produced by variations of the rules and checked for repeats using simple algorithms in \emph{Mathematica}.)

As already explained in the previous section, the first basic consequence of the rule (i) above is that any even(odd) sequence of same-side crossing terms $T_s$ can be extended repeatedly by a double $T_s$ to an arbitrary degree.  In this sense it is possible to use a compact notation for any sequence of terms in $T_s$ and $T_c$ by a prescription that indicated whether any two consecutive $T_c$s are separated by an even or an odd number of $T_s$ terms.  We utilise a notation that writes either no or one $T_s$ and implies the extension of the rule (i) summation by eventually including the factor 
\begin{equation}
1+T_s T_s + T_s T_s T_s T_s +\ldots = \left(
1-T_s T_s
\label{tssum}
\right)^{-1}.
\end{equation}
In this sense the application of rule (i) is almost trivial except when it is combined with rule (ii).  One can hence go ahead to introduce new terms by including all the possibilities for odd or even expansions of $T_s$ and then complete the series above after all other configuration types have been introduced. 

It is instructive to write down a hybrid composite of rules (i) and (ii):
\begin{itemize}
\item[(i'a)] $T_c \rightarrow T_c T_s T_c T_s T_c$
\item[(i'b)] $T_c \rightarrow T_c T_c T_s T_c T_s$
\item[(ii')] $T_s \rightarrow T_c T_s T_c$.
\end{itemize}
We note here that the two parts rule (i') can be interpreted in two ways:  either the sequence $T_s T_c T_s T_c$ is appended to the right of the original crossing $T_c$, or, the sequence $T_c T_s T_c T_s$ is added on the left of the original $T_c$.  We choose the the first of these two conventions since rules (i'a) and (i'b) produce equivalent configurations under the cyclic property -- see \ref{summationappendix}.

Consequently, the basic winding number expression can be expanded without repeating configurations under rule (i'a)
\begin{eqnarray}
Z^{(w)}_{\mathrm{basic}} &=& \left( T_c T_c \right)^w
\nonumber \\
 &\rightarrow & T_c \left(1+T_c T_s T_c T_s+T_c T_s T_c T_sT_c T_s T_c T_s+\ldots \right)\times \ldots
\nonumber \\
& = & \left( T_c \left(1-T_c T_s T_c T_s\right)^{-1} T_c \left(1-
T_c T_s T_c T_s\right)^{-1} \right)^w.
\label{Eq:PartitionFunctionLow}
\end{eqnarray}
However, this clearly does not represent a sum over all possible configurations, since rule (ii') has not been completely applied.  In principle, all configurations should be given by repeated applications of the rules to all newly introduced parts of terms.  The partition function using rule (i'a) as depicted above clearly does not repeat any configurations, yet does not produce all permissible configurations.  
We use this to calculate an \emph{approximate} partition function
\begin{equation}
Z^{(w)}_{\mathrm{appx 1}} = \left\{\frac {T_c}
{ \left(1-T_s T_s \right) \left[1-
T_s T_c \left(1-T_s T_s \right)^{-1} T_s T_c \left(1-T_s T_s \right)^{-1}\right]}\right\}^{2w}.
\label{Eq:PartFnAppx1}
\end{equation}
Integration over relevant degrees of freedom of this expression is implied. Since the weight of each configuration in eq.~(\ref{Eq:PartFnAppx1}) is the same as in the complete sum for the partition function the complete partition function for winding number $w$ given by $Z^{(w)}$ is related to the approximation as follows,
\begin{equation}
Z^{(w)}_{\mathrm{appx 1}} \leq Z^{(w)}.
\label{appx1}
\end{equation}
(We note that careful implementation of rule (ii') on a subset of $T_s$ terms above will lead to an ever better lower bound than $Z^{(w)}_{\mathrm{appx 1}}$.)

Another interpretation of iterative application of rules (i') and (ii') is given by the definition of two coupled effective terms
\begin{eqnarray}
T_c^{\mathrm{eff}} & = & T_c \left(1-T_s T_s\right)^{-1} + T_c T_s^{\mathrm{eff}} T_c T_s^{\mathrm{eff}} T_c^{\mathrm{eff}}
\label{Eq:rulesEffa}
\\
T_s^{\mathrm{eff}} & = & T_s +T_c^{\mathrm{eff}} T_s^{\mathrm{eff}} T_c^{\mathrm{eff}}.
\label{Eq:rulesEffb}
\end{eqnarray}
Here explicit evaluation shows that the scheme eventually does lead to repetition of some configurations, but all configurations are produced when combined with \eref{tssum} at the last step.  The partition function calculated using the recipe in (\ref{Eq:rulesEffa}--\ref{Eq:rulesEffb}) is defined by
\begin{equation}
Z^{(w)}_{\mathrm{appx 2}} = \left[ T^{\mathrm{eff}}_c T^{\mathrm{eff}}_c \right]^w.
\label{appx2}
\end{equation}

Now since eq.~(\ref{Eq:PartitionFunctionLow}) leads to a partition function $Z^{(w)}_{\mathrm{appx 1}}$ with correctly weighted, yet fewer configurations, and equations (\ref{Eq:rulesEffa}--\ref{Eq:rulesEffb}) yield a partition function $Z^{(w)}_{\mathrm{appx 2}}$ with all yet some multiply occurring configurations we know how the true partition function is bounded
\begin{equation}
Z^{(w)}_{\mathrm{appx 1}} \leq Z^{(w)} \leq Z^{(w)}_{\mathrm{appx 2}}.
\end{equation}
In principle these two approximations are calculable in the scenario of a polymer loop winding around certain obstacles in the plane, as described in sections \ref{slitsection} and \ref{generalsection}, and can be used to understand upper and lower bounds for free energy associated with a particular winding number. We calculate only the lower bound $Z^{(w)}_{\mathrm{appx 1}}$ here, as the nonlinear coupled equations \eref{Eq:rulesEffa} and \eref{Eq:rulesEffb} pose formidable challenges. As stated, the approximations still need to be integrated over the relevant degrees of freedom, as will be set out in section \ref{diagsect}.

\subsection{Counting the number of crossing or same-side terms}
\label{countcrossings}
In either of the suggested approximations for the partition function (see \eref{Eq:PartFnAppx1} and \eref{appx2}) it is possible to include generating terms that may be used to calculate the number of $T_s$ or $T_c$ terms. If in these summations we simply replace $T_s \rightarrow e^{g_s}T_s$ and $T_c \rightarrow e^{g_c}T_c$, then we may calculate the average number of crossing terms as
\beq
\langle N_c \rangle = \left[\frac{\partial}{\partial g_c} \log Z^{(w)}(g_c,g_s) \right]_{g_c = g_s = 0},
\label{ncross}
\eeq
and the average number of same-side terms as
\beq
\langle N_s \rangle = \left[\frac{\partial}{\partial g_s} \log Z^{(w)}(g_c,g_s) \right]_{g_c = g_s = 0}.
\label{nsame}
\eeq

\subsection{Probability distribution of a flexible polymer in half-space}
\label{tdetails}
We still have to assign a statistical weight to each string in the summation over all diagrams. We proceed to do this for a flexible polymer. In principle other polymer variants could be described by the formalism up to this point, but the form of the probability distribution would be different.

For a flexible polymer we treat the sub-arcs (labelled by the various $T$s) as random walks confined to half-spaces. In polymeric systems with suitable solubility and flexibility conditions this is, of course, a reasonable assumption \cite{edwards1975size}. The notion of confined random walks is certainly not a new one. In a 1943 review Chandrasekhar \cite{chandrasekhar} pointed out how one-dimensional random walks with reflecting and absorbing boundary conditions may be treated. Naturally boundaries change the probability distribution for random walks. A reflection off such a boundary implies that a walker must necessarily retrace its last step. An absorbing boundary, in contrast, would prevent any further displacements. Consequently the probability distribution of a walker confined by a reflecting boundary is obtained by adding an ``image distribution'' to that of an unconfined walker. This accounts for additional possible paths to a given end point, stemming from the reflecting boundary (these paths may be viewed as mirrored paths in the excluded region). On the other hand, the distribution of a walker confined by an absorbing boundary is obtained by subtracting a similar mirror distribution. This, in turn, accounts for the exclusion of trajectories that terminate on the absorbing boundary.

This discussion may be extended to a random walker restricted to a half-space in three dimensions (see, for instance, the article of Slutsky \cite{slutsky} where a similar ``method of images'' is used). We shall draw on these notions in order to assign appropriate statistical weights to the sub-arcs mentioned in the previous section. To this end we make the following assumptions (as illustrated in \Fref{fig:epsilon}):
{\begin{enumerate}
	\item there exists a finite minimal length scale (such as a bond length or Kuhn length) in this polymer system,
	\item for every (sub)sequence $T_{xy}T_{yz}$ (for any $x,y,z\in\{+,-\}$), there exists a trans-plane polymer segment of length $2\epsilon$ that is normal to the plane and connects the two sub-arcs between the $z>0$ or the $z<0$ half-spaces, \label{epsilonassupt1}
	\item thus any given sub-arc begins and ends at a distance $\epsilon$ from the $xy$-plane (see \Fref{fig:epsilon}),\label{epsilonassupt2}
	\item each polymer sub-arc is modelled as a random walk constrained by an absorbing boundary plane to either the $z>0$ or the $z<0$ half-spaces,
	\item such a random walk begins and ends at $\vec r_0 = (x_0,y_0,\eta\epsilon)$ and $\vec r = (x,y,\eta\epsilon)$ respectively (here $\eta =+ 1$ for the $z>0$ half-space or $\eta=-1$ for $z<0$),
	\item the random walks are fully flexible, and each has a variable arc-length $s_i$.
\end{enumerate}
The assumption of an absorbing boundary is based on the fact that we are interested in the two piercings that a sub-arc makes with the plane, since it is there that one particular sub-arc ends and another one begins.
\begin{figure}[h!]
        \centering
        \begin{subfigure}[b]{0.4\textwidth}
                \centering
								\includegraphics[width=0.9\textwidth]{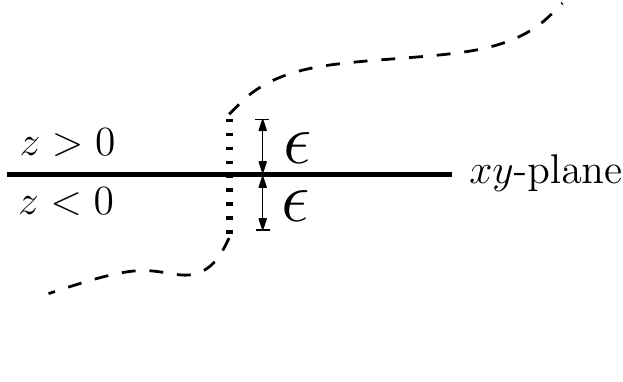}
								\caption{Side view of trans-plane segment.}
								\label{fig:transplane}
        \end{subfigure} \quad
        \begin{subfigure}[b]{0.565\textwidth}
                \centering
								\includegraphics[width=0.9\textwidth]{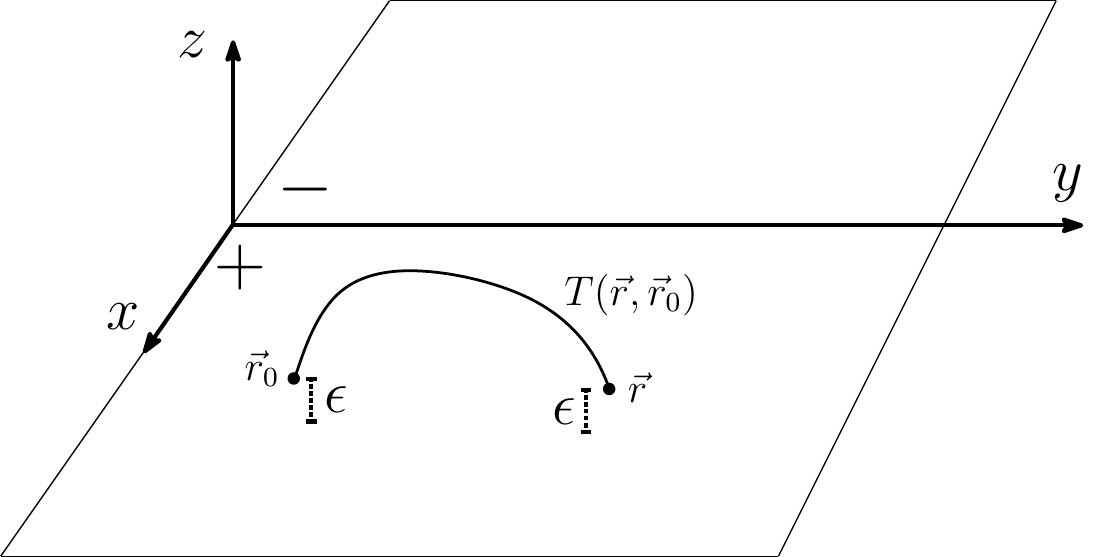}
	  						\caption{Polymer in the half plane $z>0$.}
								\label{fig:basics2}
        \end{subfigure}%
       	\caption{Sub-arcs as random walks that begin and end at a distance $\epsilon$ from the plane.}
        \label{fig:epsilon}
\end{figure}
In reality, the trans-plane connecting segment should be free to take on any orientation. Our approximation that it is normal to the plane should be a small correction for a sufficiently long arc segment. It is clear from \Fref{fig:epsilon} that, barring constraints due to the total length of the polymer, the $y$ components of the beginning and end vectors of a random walk could take any value. The $x$ components, however, are restricted either to the interval $(0,\infty)$ or to the interval $(0,-\infty)$, depending on whether they fall on the $+$ or $-$ side of the rod. Naturally this will constrain the integration bounds for $x$ in the partition function accordingly.

Therefore, for a Gaussian chain with all lengths expressed in terms of the Kuhn length, we may now assign the corresponding probability distribution to a typical sub-arc as set out in \cite{slutsky},
\begin{eqnarray}
T(\vec r,\vec r_0,s,\epsilon) &= (2\pi s)^{-3/2} \left[ e^{-(\vec r - \vec r_0)^2/2s}-e^{-(\vec r + \vec r_0)^2/2s} \right] \nl
&=  (2\pi s)^{-3/2}  e^{-[(x-x_0)^2+(y-y_0)^2]/2s}\left[1-e^{-\epsilon^2/s}\right],
\label{prob1}
\end{eqnarray}
where $s$ is the arc length of the polymer sub-arc between $\vec r_0$ and $\vec r$. The $z$ dependence vanishes due to assumptions (\ref{epsilonassupt1}) and (\ref{epsilonassupt2}) above. Here we have interchanged the discrete number of steps $N$ for the arc-length variable $s$ through appropriate re-scaling. Finally, given a polymeric system with a sufficiently small Kuhn length, we may assume $\epsilon \ll 1$ and Taylor expand the exponential in \Eref{prob1} to obtain
\beq
T_p(\vec r,\vec r_0,s,\epsilon) = \frac{\epsilon^2}{\sqrt{(2\pi)^{3}s^{5}}}  e^{-[(x-x_0)^2+(y-y_0)^2]/2s}.
\label{tp}
\eeq
The subscript $p$ here simply refers to the ``parity'' of the particular $T$ under consideration, as described in \sref{condensednotation}. This merely indicates whether the $x$ co-ordinates are on the same side of the rod or not, which will determine the integration domains for the $x$ co-ordinates in the partition function.


\subsection{Partition function for $w=1$}
\label{diagsect}
Using \sref{rules} we constructed a symbolic summation of all possible configurations for $w=1$ in two possible approximations, \eref{Eq:PartFnAppx1} and \eref{appx2}. We now proceed to use this summation in order to write the partition function for this system. Let us denote the set of all valid configurations as $\Lambda$. Supposing that the total length of the polymer loop is $L$, we note that
\beq
\mathbf Z^{(w=1)}_{\mathrm{total}}(L) = \sum_{\chi \in \Lambda} Z_\chi^{(w=1)}(L),
\eeq
i.e., the total partition function is simply the sum of the partition functions for all valid configurations for $w=1$. A typical valid configuration $\chi$ which has $N$ piercings of the $xy$-plane is represented by some sequence (of length $N$) of $T$s that adheres to the conditions in \Sref{validseq}, and simply corresponds to one of the terms in the summation. Let us make the notation in equations \eref{zfig1} and \eref{zfig2} more concrete: the partition function corresponding such a generic sequence would be
\beqa
Z^{(w=1)}_\chi(L) &=& \int_0^\infty dX \int_{-\infty}^\infty dY \int_\epsilon^\infty dS \;\;\delta(x_0-x_N)\delta(y_0-y_N) \nl
&&\;\;\;\prod_{j=1}^N T_{p_j}(x_{j-1},x_j;y_{j-1}-y_j;s_j;\epsilon)\delta(\sum_{k=1}^N s_k-L).
\label{zgeneral}
\eeqa
The condensed notation implies
\beqa
dX &=& dx_0\, dx_1\ldots dx_N, \nl
dY &=& dy_0\, dy_1\ldots dy_N, \nl
dS &=& ds_1\, ds_2\ldots ds_N,
\eeqa
and the functions $T_p$ in the integrand are each of the form \eref{tp}. Naturally there is one less $s$ integral than for $x$ or $y$, since one arc-segment connects two planar points. The first two delta functions ensure that the strand is closed: the first and last $x$ and $y$ co-ordinates must be equal. The length $s_i$ of each of the $N$ sub-arcs is bounded from below by the minimal length $\epsilon$ - hence the integration bounds on the $s$ integrals. It is, however, necessary that these lengths add up to the total length $L$ of the entire polymer. This constraint is enforced by the third delta function. As set out in \Sref{tdetails}, the $y$ co-ordinates of each strand are not constrained. For this reason they are integrated over the whole axis. From \eref{prob1} it is clear that the $T$s are symmetric under the exchanges $(x_{i-1}-x_i)\rightarrow-(x_{i-1}-x_i)$ and $(y_{i-1}-y_i)\rightarrow-(y_{i-1}-y_i)$. Since the $x$ co-ordinates are constrained to one half of their axis (as set out after \Eref{tp}), we need to distinguish between the terms that cross over the rod ($T_c$) and those that remain on the same side of the rod ($T_s$). For $T_s$, the two $x$ argument have the same sign and the function depends on $\pm(x_{i-1}-x_i)$. For $T_c$, the two $x$ arguments have opposite signs and the function depends on $\pm(x_{i-1}+x_i)$. We may thus change all $x$ integrals to run over $(0,\infty)$ under the condition that
\beqa
T_s(x_{i-1},x_i)&=&T(x_{i-1}-x_i)=\frac{\epsilon^2}{\sqrt{(2\pi)^{3}s_i^{5}}}  e^{-[(x_{i-1}-x_i)^2+(y_i-y{i-1})^2]/2s_i},\nl
T_c(x_{i-1},x_i)&=&T(x_{i-1}+x_i)=\frac{\epsilon^2}{\sqrt{(2\pi)^{3}s_i^{5}}}  e^{-[(x_{i-1}+x_i)^2+(y_i-y{i-1})^2]/2s_i},
\label{tsandtc}
\eeqa
i.e., with these integration bounds the same-side contributions depend on the difference between their $x$ co-ordinates, whereas the crossing contributions depend on the sum (compare to \eref{tp}). For the reasons set out above, it is clear that translational invariance holds for the $y$ co-ordinates but not for the $x$ co-ordinates. In \ref{diagappend} we outline how some integrals in (\ref{zgeneral}) may be diagonalised using a Laplace transformation in the length of the polymer and Fourier transformations of the $y$ co-ordinates. The result is the Laplace transformed partition function for some configuration $\chi$,
\beq
\tilde Z^{(w=1)}_\chi (t) = \int_0^\infty dX \;\delta(x_0-x_N)\int_{-\infty}^\infty dk\;\; \prod_{i=1}^N  T_{p_i}^{L,F}(x_{i-1},x_i;k;t),
\label{ztilde}
\eeq
where the superscript ``$L,F$'' implies that the $T$s from \eref{tsandtc} have been Laplace transformed and Fourier transformed in $y$. We have thus obtained a diagonalisation for the $s$ and $y$ co-ordinates. What remains are the integrals over the positive (real) $x$ axis, and the integral over the Fourier variable $k$. We shall deal with these integrals for two cases. First we consider restricting the $x$ integrals to a narrow slit, thereby effectively constraining the polymer to be wound through two slits in the plane. In this case no $x$ integration is necessary, and only the $k$ integral remains. Secondly we shall outline possible approximation schemes to deal with the general case.

As stated, the complete partition function $Z^{(w=1)}$ is found by the summation over various diagrams. This summation is approximated by \eref{Eq:PartFnAppx1} or \eref{appx2}. The integrand in this partition function may then be repeated $w$ times to obtain the approximated partition function for higher winding numbers $w>1$.

\section{Specific case: polymer wound through two slits}
\label{slitsection}
Let us consider a polymer looping around two slits, the inner edges of the slits separated by the distance $d$ and each slit with a width $\Delta$, as shown in Figure \ref{fig:doubleslit}. We note that similar scenarios of confined wound polymers have been considered in \cite{grosbergfrish2003}.
\begin{figure}[H]
	\centering
		\includegraphics[width=0.60\textwidth]{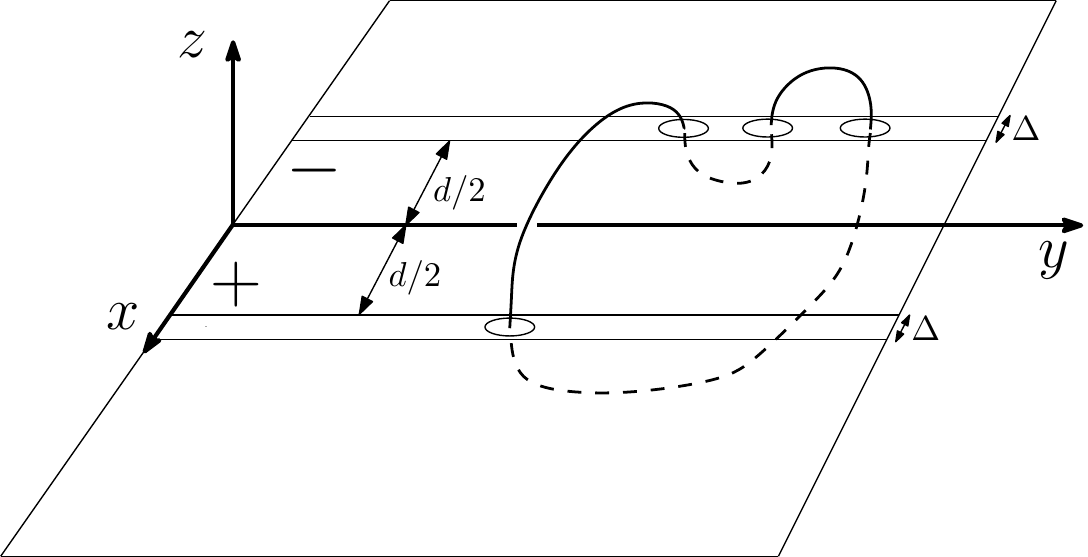}
	\caption{Constraining the polymer to two narrow slits in the plane.}
	\label{fig:doubleslit}
\end{figure}  
We can compute the average number of the types of arc elements in this scenario using the formalism developed in \sref{countcrossings}. By winding the polymer around a double slit geometry, the combinatorics clearly remain unaltered.  The partition function for the chain now has integration restricted over the domain $x_i\in [d/2,d/2+\Delta]$.  This means that equation \eref{ztilde} takes the integral form
\beq
\tilde{Z}^{(w=1)}_\chi (t,\Delta,d) = \int_{\frac d 2}^{\frac d 2 +\Delta}dX\;\delta\left(x_0-x_N\right) \int_{-\infty}^\infty dk\; \prod_{i=1}^N  T_{p_i}^{L,F}(x_{i-1},x_i;k;t).
\label{Eq:PartitionFn-wd}
\eeq
We begin by considering the case of zero slit-width.

\subsection{Zero slit width: $\Delta = 0$}
\label{deltazerosection}
In the limiting case where $\Delta\rightarrow 0$ such that each ${T}$ has exactly the same $x$--value $x_i=d/2,\,\,\forall i\in \{0,\ldots,N\}$, the partition function becomes especially simple as the operator sums of eq.~(\ref{Eq:PartFnAppx1}) or (\ref{appx2}) now become  simple algebraic sums. (This is the scenario where no integration is necessary, and conformations are simply summed because the problem is ``diagonal'' already in the simplest terms.  The next section will deal with the extension of this to narrow and easily integrable slits.) Choosing the first approximation (\ref{Eq:PartFnAppx1}) we obtain
\beqa
\tilde Z^{(w=1)}_{\mathrm{appx 1}}(t,d) &=& \int_{-\infty}^\infty dk \;\;\frac{ (T_c^{L,F})^2 [ 1- (T_s^{L,F})^2]^2}
{\left[1-[2+  (T_c^{L,F})^2] (T_s^{L,F})^2+ (T_s^{L,F})^4\right]^2},
   \label{ztildedelta0}
\eeqa
since the operations of summing over various configurations and integration over $k$ may be exchanged. 

Turning to equation \eref{tsandtc} we see that Fourier transformation in $y$ leads us to the following two cases,
\begin{equation}
 T^{F}_s(d;k;s_i)=\frac{\sqrt{3} \epsilon ^2 e^{-\frac{k^2 s_i}{6}}}{2 s_i^2}
\label{delta0a}
\end{equation}
and
\begin{equation}
 T_c^{F}(d;k;s_i)=\frac{\sqrt{3} \epsilon ^2 e^{-\frac{3 d^2}{2 s_i}-\frac{k^2 s_i}{6}}}{2 s_i^2}.
\label{delta0b}
\end{equation}
Since $\epsilon$ is non-zero and finite, these expressions are well-defined. We now need to perform the Laplace transforms of each of these,
\beqa
 T_s^{L,F}(d;k;t) &=& \int_\epsilon^\infty ds_i \;\; e^{-s_i t} \frac{\sqrt{3} \epsilon ^2 e^{-\frac{3 d^2}{2 s_i}-\frac{k^2 s_i}{6}}}{2 s_i^2} \nl
&=& \frac{\sqrt{3} \epsilon ^2}{2}  \left[\frac{e^{-\frac{1}{6} \epsilon \left(k^2+6
   t\right)}}{\epsilon}-\frac{1}{6} \left(k^2+6 t\right) \Gamma
   \left(0,\frac{1}{6} \left(k^2+6 t\right) \epsilon \right)\right]
\label{delta0c}
\eeqa
and
\beqa
 T_c^{L,F}(d;k;t) &\approx& \int_{d^2}^\infty ds_i \;\; e^{-s_i t} \frac{\sqrt{3} \epsilon ^2 e^{-\frac{3 d^2}{2 s_i}-\frac{k^2 s_i}{6}}}{2 s_i^2}.
\label{delta0d}
\eeqa
In \eref{delta0c} the answer contains an incomplete gamma function. The integral in \eref{delta0d} has an approximated lower bound of $d^2$ (recall that $s_i$ and $d$ are dimensionless). In principle this bound should be $\epsilon$ (i.e., of the order of the Kuhn length). If the polymer were inextensible, the minimum arc-length for $T_c^{L,F}$ should be $d$. Although we deal with a Gaussian chain here, this approximation is reasonable since the integrand is dominated by $s_i\geq d^2$. We approximate
\beqa
T_c^{L,F}(d;k;t) &\approx& \frac 1 2 e^{-\frac{3 d^2}{2 s^*}}\,\int_{d^2}^\infty ds_i \;\; e^{-s_i t} \frac{\sqrt{3} \epsilon ^2 e^{-\frac{k^2 s_i}{6}}}{2 s_i^2},
\label{delta0e}
\eeqa
where $s^* = (\int_{d^2}^\infty ds \;\; s\, e^{-\frac{k^2 s}{6}})/ (\int_{d^2}^\infty ds \;\; e^{-\frac{k^2 s}{6}}) =  \frac{6+ d^2 k^2}{k^2}$ is a constant value that captures some of the $k$ scaling of the answer. Numerical verification shows this approximation to perform very well for various ranges of $t$ and $k$. The integral in \eref{delta0e} may now be evaluated explicitly,
\beqa
T_c^{L,F}(d;k;t) &=& \frac{\sqrt{3} \epsilon ^2}{2}  e^{-\frac{3 d^2 k^2}{2\left(6+d^2
   k^2\right)}}  \nl
&& \times \left[\frac{e^{-\frac{1}{6} d^2 \left(k^2+6
   t\right)}}{d^2}-\frac{1}{6} \left(k^2+6 t\right) \Gamma
   \left(0,\frac{1}{6} d^2 \left(k^2+6 t\right)\right)\right].
\label{delta0f}
\eeqa

We may now insert the answers \eref{delta0c} and \eref{delta0f} into \eref{ztildedelta0} to obtain two approximations for the partition function. Naturally the integrand above is some very complicated function of $k$. As is verifiable numerically, however, a saddle point approximation is reliable for various ranges of $t$ and $d$. We omit the cumbersome explicit form of the result.

The average length of the loop for $w=1$ may be calculated from \eref{ztildedelta0},
\beq
\langle L \rangle (t) = -\frac{\partial}{\partial t} \log \left[\tilde Z^{(w=1)}_{\mathrm{appx 1}}(t,d)\right].
\label{laveeq}
\eeq
In Figure \ref{fig:Lave} we see that $\langle L \rangle (t)$ is a concave function for various values of the slit separation $d$. Here we have set $\epsilon=1$ for convenience; this convention is used henceforth. It is also clear that small Laplace parameters and longer length scales correspond, particularly for small $d$. As we increase the size of $d$, we note that $\langle L \rangle$ seems to strive asymptotically to increasingly large values. This is simply a manifestation of the fact that the non-zero slit-separation implies a minimal length scale for the polymer loop. 

We thus have a ``lookup table'' that allows us to associate an average length of the polymer to a particular Laplace parameter (of course this is not the inverse Laplace transform, as would ideally be the case). 
\begin{figure}[H]
	\centering
		\includegraphics[width=0.60\textwidth]{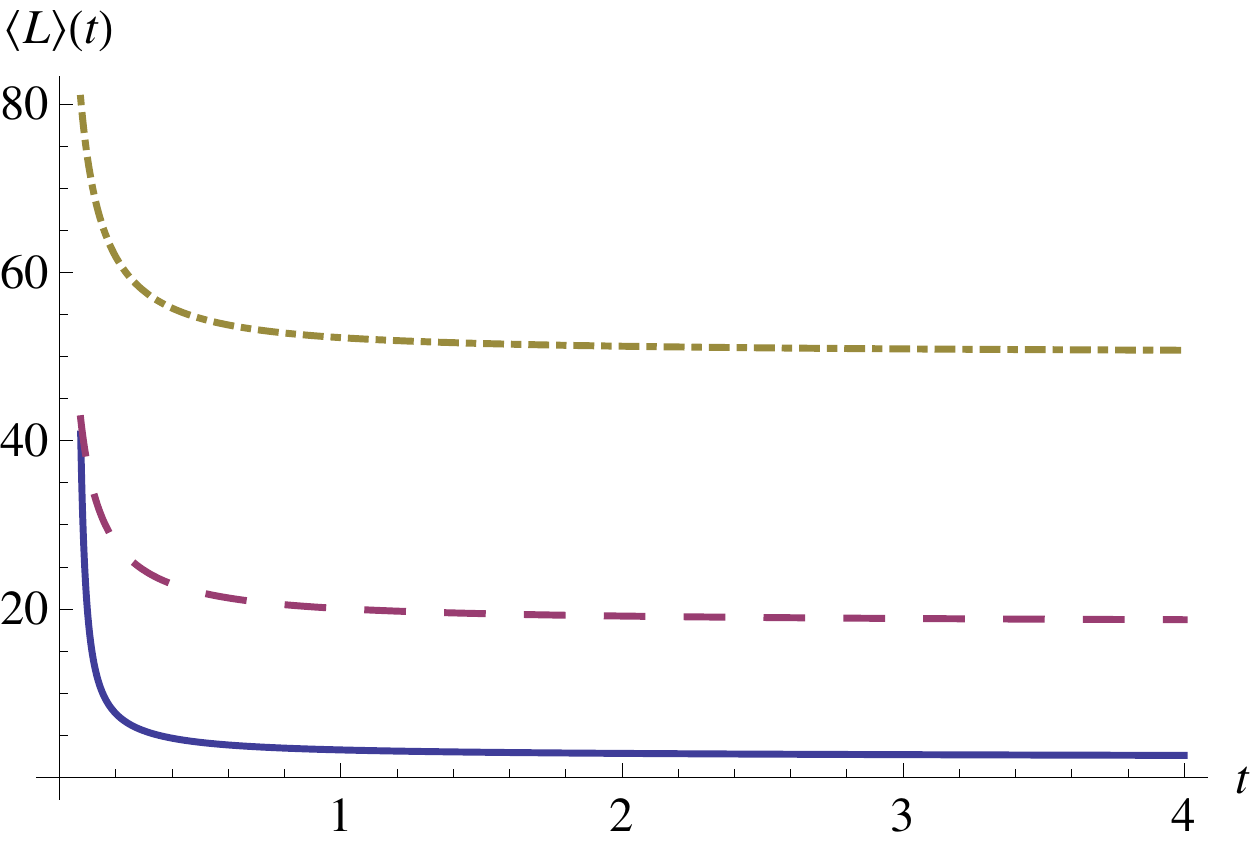}
	\caption{Average length of the loop as function of the Laplace parameter $t$, calculated according to equation \eref{laveeq}. Parameters: $d=1$ (solid), $d=3$ (dashed), $d=5$ (dashdotted).}
	\label{fig:Lave}
\end{figure}

It is further interesting to ask what is the relative weight of the undressed term / basic loop $T_c T_c$ (see Figure \ref{fig:fig1}) in the summation over all valid diagrams for $w=1$. To this end we simply look at the probability for this configuration,
\beq
P(T_c^2) = \frac{\int_{-\infty}^\infty dk \;\;  (T_c^{L,F})^2}{\tilde Z^{(w=1)}_{\mathrm{appx 1}}(t,d)}.\label{ptc}
\eeq
We see in Figure \ref{fig:prob} that, for various values of $d$, this probability is unity for sufficiently large Laplace parameters. Through Figure \ref{fig:Lave} we may thus identify length-scales (for various slit separations) at which the basic (undressed) loop provides the dominant contribution to the partition function. This makes sense physically, since large Laplace parameters correspond to short length-scales. Naturally the afore-mentioned minimal length scale set by the slit separation implies that as soon as the length of the loop becomes small enough, the basic configuration consisting of two crossing terms will be the dominant configuration in the partition function.
\begin{figure}[H]
	\centering
		\includegraphics[width=0.60\textwidth]{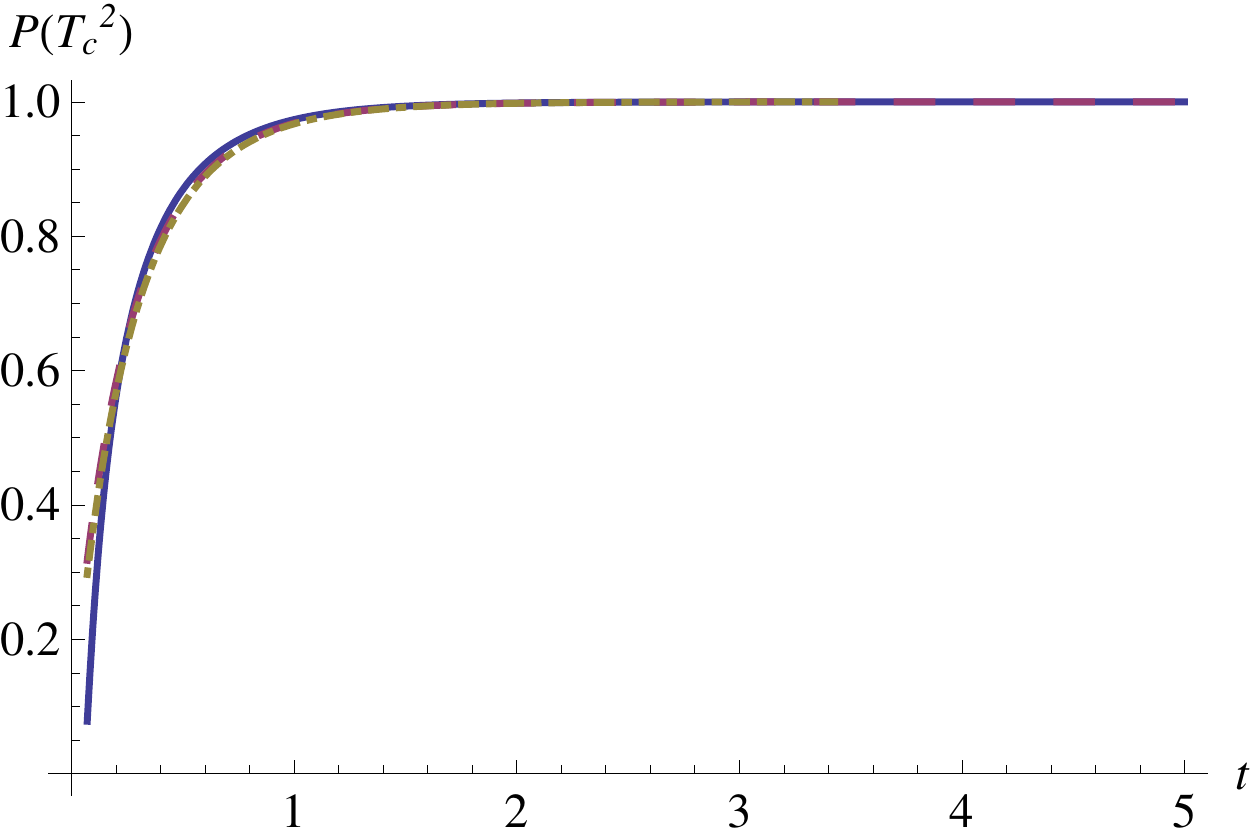}
	\caption{Probability for the configuration $T_c^2$ as a function of the Laplace parameter $t$ (calculated according to equation \eref{ptc}). Parameters: $d=1$ (solid), $d=3$ (dashed), $d=10$ (dashdotted).}
	\label{fig:prob}
\end{figure}

The average number of crossing terms and same-side terms may also be calculated as a function of the Laplace parameter according to equations \eref{ncross} and \eref{nsame}. From Figure \ref{fig:aveN} we note that as soon as the Laplace parameter is sufficiently large (i.e., the polymer is typically short), there are exactly two $T_c$ terms and zero $T_s$ terms present. This agrees with the previous conclusion: at short polymer lengths, the undressed basic term $T_c T_c$ dominates the summation over valid diagrams. We note that as the slit separation $d$ is increased, this undressed configuration dominates at decreasing $t$, i.e., at longer polymer lengths. This can be related to Figure \ref{fig:Lave}, where we observe the minimal length of the polymer increasing with increases in slit separation.

As is to be expected for sufficiently small polymer length-scales, the results are not particularly sensitive to which approximation is used to approximate the summation. Indeed, for short or stiff polymers it should be sufficient to generate the first few terms (valid sequences) explicitly as an approximation to the complete partition function.

\begin{figure}[h]
        \centering
        \begin{subfigure}[b]{0.48\textwidth}
                \centering
								\includegraphics[width=\textwidth]{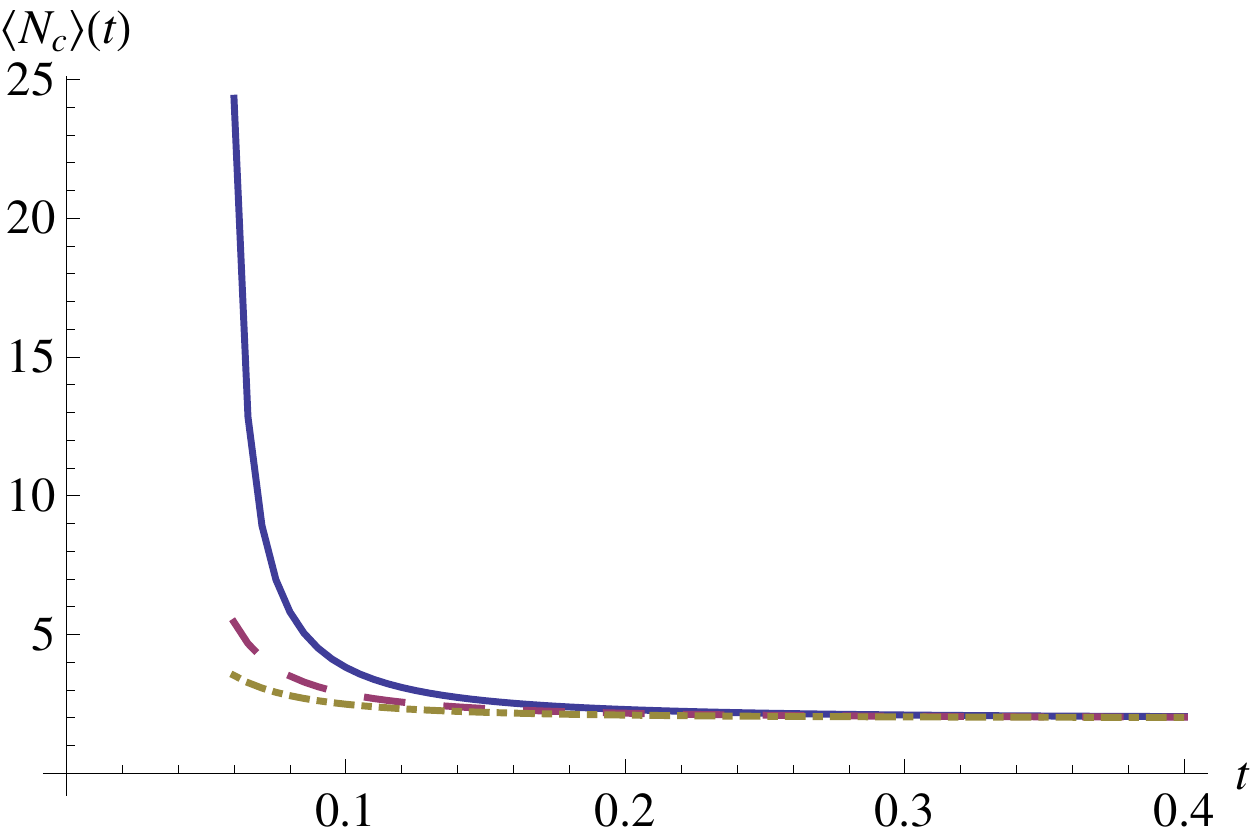}
								\caption{Average number of crossing terms}
								\label{fig:nc}
        \end{subfigure} \quad
        \begin{subfigure}[b]{0.48\textwidth}
                \centering
								\includegraphics[width=\textwidth]{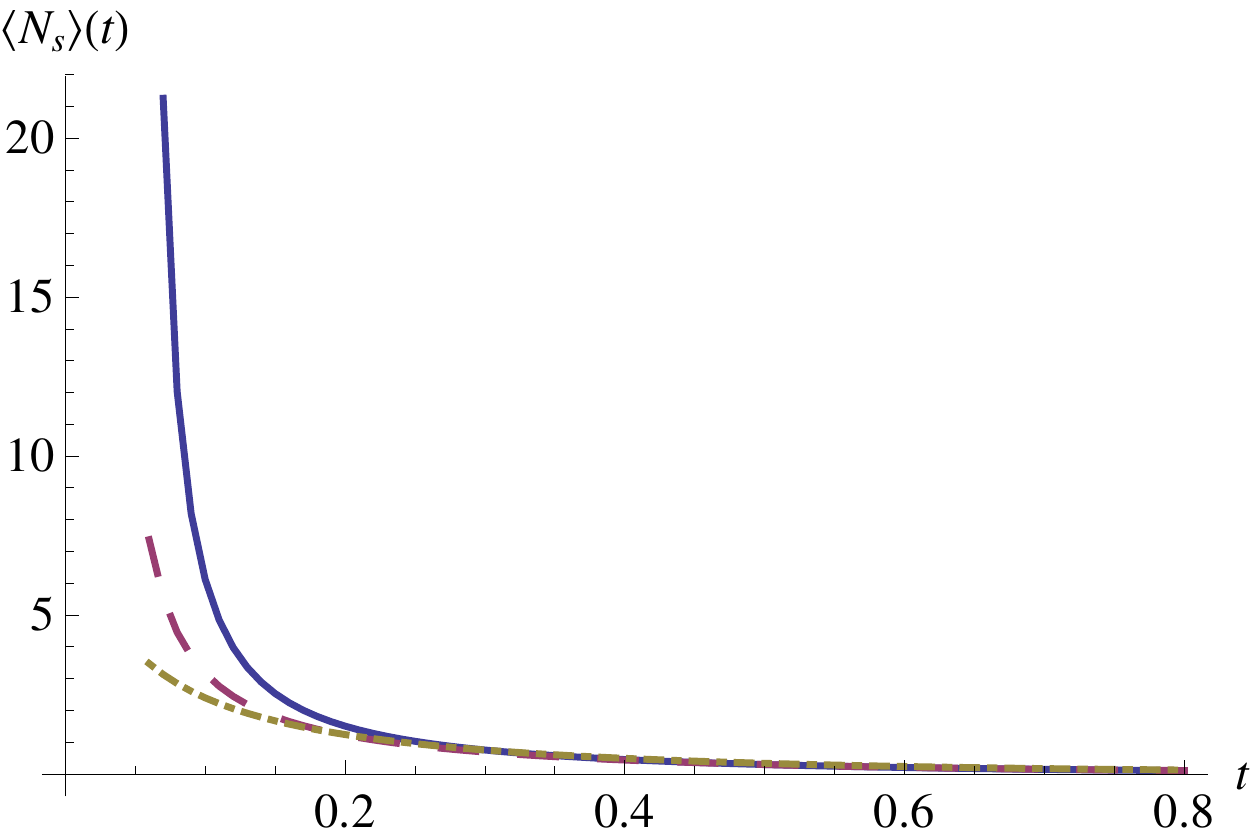}
	  						\caption{Average number of same-side terms}
								\label{fig:ns}
        \end{subfigure}%
       	\caption{The average number of crossing and same-side terms as functions of Laplace parameter, calculated numerically from \eref{ncross} and \eref{nsame}. Parameters: $d=1$ (solid), $d=2$ (dashed), $d=3$ (dashdotted).}
        \label{fig:aveN}
\end{figure}

Lastly we consider the free energy of the system for various values of slit separation. This quantity is simply the negative logarithm of the partition function, and is plotted parametrically as a function of the average polymer length in Figure \ref{fig:Fdeltazero}. It is clear that as the polymer is made shorter and approaches the minimal length-scale set by the slit separation, the free energy increases sharply which is compatible with the sharp decrease of entropy experienced by the Gaussian chain. 
\begin{figure}[H]
	\centering
		\includegraphics[width=0.60\textwidth]{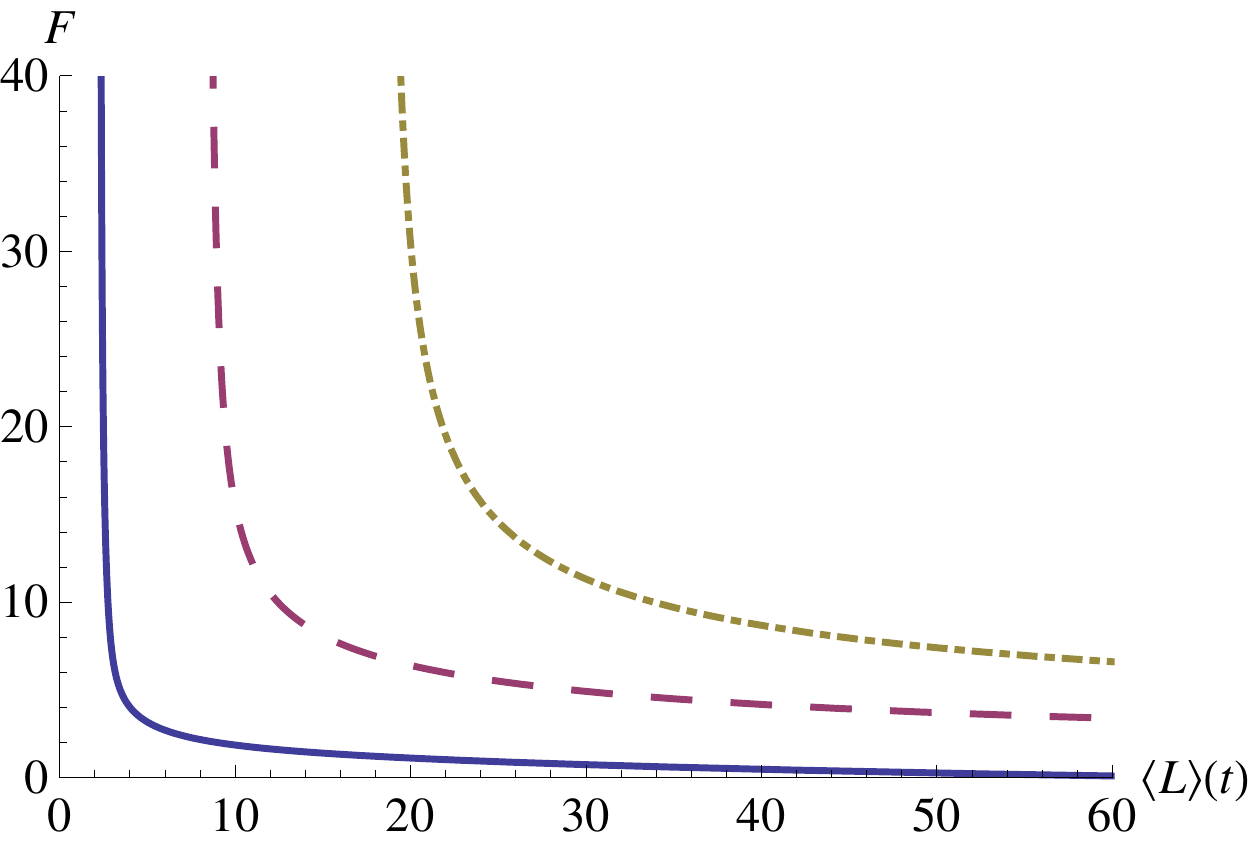}
	\caption{Parametric plot of free energy dependence on average polymer length. Parameters: $d=1$ (solid), $d=2$ (dashed), $d=3$ (dashdotted).}
	\label{fig:Fdeltazero}
\end{figure}

\subsection{Finite slit width: $\Delta \neq 0$}
We shall briefly outline a possible approach to solving the case for non-zero slit width in \eref{Eq:PartitionFn-wd}. As a first order approach it would be sensible to decouple consecutive $T$s completely and to replace the various $x$ integrals ($\int_{\frac d 2}^{\frac d 2 +\Delta}\;dx_i$) with a localisation approximation,
\beq
\int_{-\infty}^{\infty}dx\;\int_{-\infty}^{\infty} dx'\; \frac {\mathcal N} {\Delta} \;\;\exp\left[{-\frac{(x-\frac d 2)^2}{\Delta}-\frac{(x'-\frac d 2)^2}{\Delta}-\frac{(x-x')^2}{s}}\right]
\eeq
and
\beq
\int_{-\infty}^{\infty}dx\;\int_{-\infty}^{\infty} dx'\; \frac {\mathcal N} {\Delta} \;\;\exp\left[{-\frac{(x-\frac d 2)^2}{\Delta}-\frac{(x'-\frac d 2)^2}{\Delta}-\frac{(x+x')^2}{s}}\right]
\eeq
for $T_s$ and $T_c$ respectively.  The analogues of equations \eref{delta0a} and \eref{delta0b} now become
\beq
T_s^{F}(d;k;s_i)=  \frac{\sqrt{3} \epsilon ^2 e^{-\frac{k^2 s_i}{6}}}{2 s_i^2} \sqrt{\frac{s_i}{s_i+ 2 \Delta}}
\label{nonzerodelta1}
\eeq
and
\beq
T_c^{F}(d;k;s_i)=\frac{\sqrt{3} \epsilon ^2 e^{-\frac{3 d^2}{2 s_i}-\frac{k^2 s_i}{6}}}{2 s_i^2} \sqrt{\frac{s_i}{s_i+ 2 \Delta}} e^{-\frac{3 d^2}{2 \left(2 \Delta ^2+s_i\right)}}.
\label{nonzerodelta2}
\eeq
Naturally these expressions reduce to the case of zero slit-width if $\Delta \to 0$. Again we require Laplace transformations in order to use \eref{Eq:PartFnAppx1}, and we must do the $k$ integral as for \eref{ztildedelta0}. This may be approximated in various ways. A Taylor expansion of \eref{nonzerodelta1} and \eref{nonzerodelta2} to $\mathcal O (\Delta)$, for instance, allows us to repeat the analysis from \sref{deltazerosection} without many modifications, yielding an approximation for Laplace transformed partition function, $\tilde Z^{(w=1)}_{\mathrm{appx 1}}(t,d,\Delta)$. (We may recover \eref{ztildedelta0} through $\lim_{\Delta \to 0} \tilde Z^{(w=1)}_{\mathrm{appx 1}}(t,d,\Delta) = \tilde Z^{(w=1)}_{\mathrm{appx 1}}(t,d)$.) One may now calculate, for instance, the force exerted by the slit as the derivative of the free energy,
\beq
f(t,d,\Delta)= -\frac\partial{\partial \Delta} \left(-\log \left[\tilde Z^{(w=1)}_{\mathrm{appx 1}}(t,d,\Delta)\right]\right).
\eeq
With the aid of \emph{Mathematica} we may ``invert'' the Laplace transformation numerically to obtain the Laplace parameter as a function of the average length, $t = t(\langle L \rangle, d,\Delta)$. This allows us to plot, for instance, the force as a function of slit separation for a fixed $\langle L \rangle$, see Figure \ref{fig:forceratio}.

\begin{figure}[H]
        \centering
        \begin{subfigure}[b]{0.48\textwidth}
                \centering
								\includegraphics[width=\textwidth]{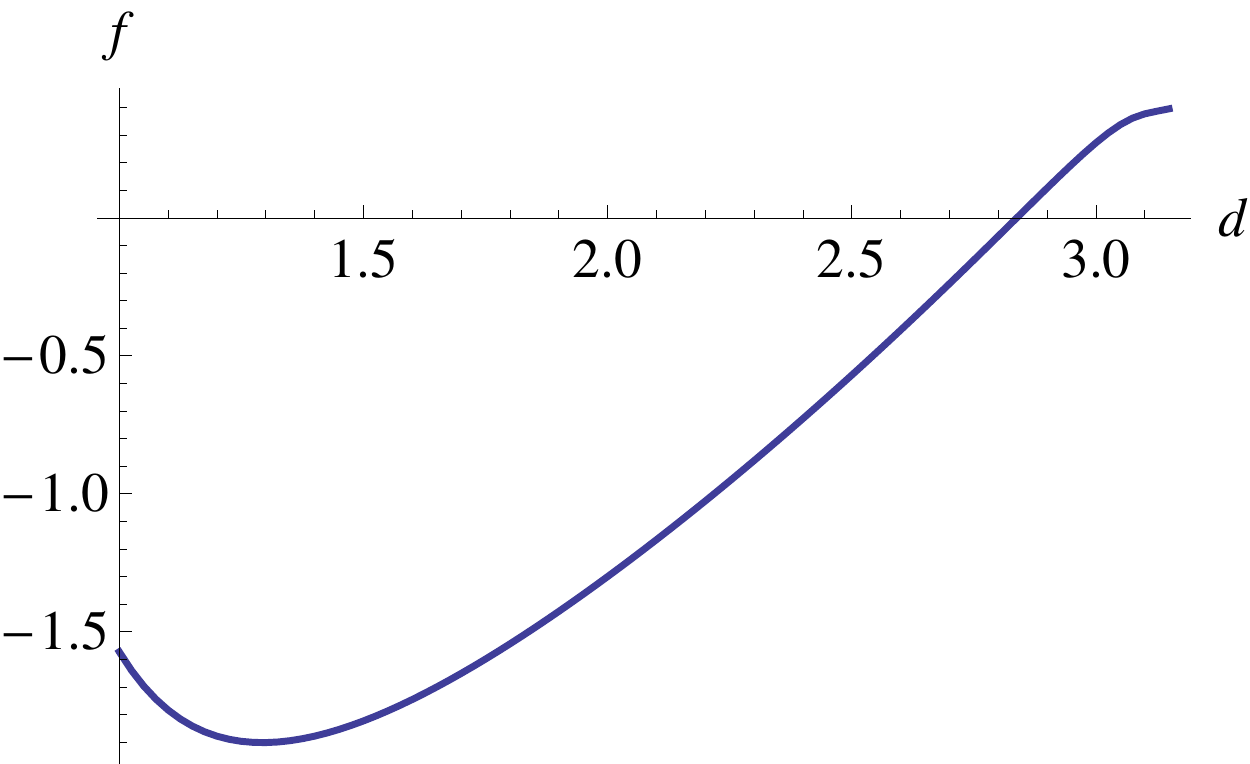}
								\caption{Force dependence on $d$.}
								\label{fig:force}
        \end{subfigure} \quad
        \begin{subfigure}[b]{0.48\textwidth}
                \centering
								\includegraphics[width=\textwidth]{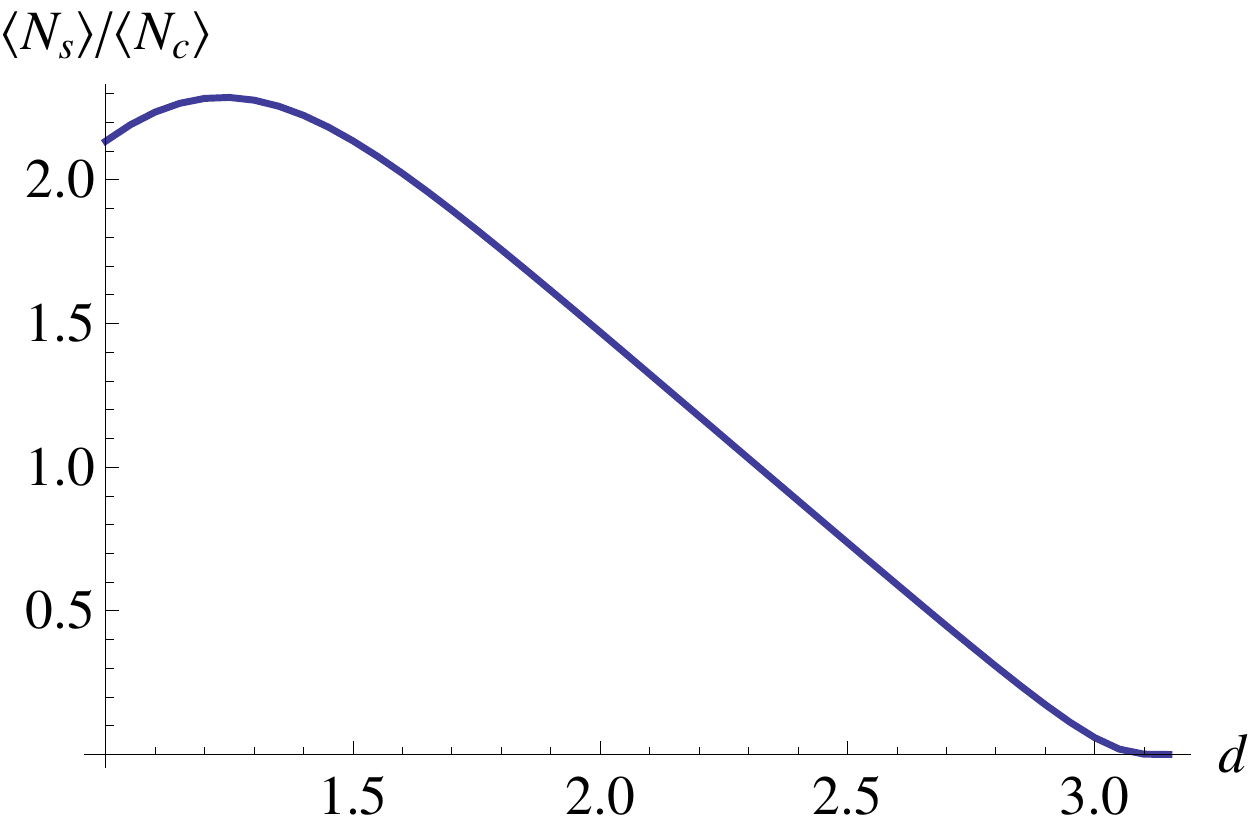}
	  						\caption{$\frac{\langle N_s \rangle}{\langle N_c \rangle}$ as function of $d$.}
								\label{fig:ratio}
        \end{subfigure}%
       	\caption{Force exerted by the slit as a function of slit separation $d$, for $\Delta = 0$ and a fixed $\langle L \rangle=20$. The ratio $\frac{\langle N_s \rangle}{\langle N_c \rangle}$ exhibits a peak corresponding to the minimum of the force. Compare to Figure \ref{fig:Lave} to see why $d>3.1$ is excluded.}
        \label{fig:forceratio}
\end{figure}
  
Despite the somewhat crude simplifying assumptions made (using independent localisation of each arc), expected physical aspects are well captured. For a fixed average polymer length, there is a sign change in the force at some slit separation. This makes sense: for sufficiently small $d$ the slit has a ``compressing'' effect on the polymer, and for sufficiently large $d$ the polymer is ``stretched''. Again the competition of the two length scales $d$ and $\langle L \rangle$ is manifest. In Figure \ref{fig:ratio} we plot the ratio $\frac{\langle N_s \rangle}{\langle N_c \rangle}$ as defined in \eref{ncross} and \eref{nsame} for a fixed $\langle L \rangle$. Consider the trend in this plot as we decrease the slit-separation $d$. For large $d$ all of the polymer is in the crossing terms and $\langle N_s \rangle \approx 0$. As we decrease $d$ the fraction of same-side terms increases: more of the polymer's length is free to occupy the slits. As $d$ is decreased even further, the ratio begins decreasing again: at some stage sufficiently much polymer length is free so that additional crossing terms may arise, thereby decreasing $\langle N_s \rangle$ and increasing $\langle N_c \rangle$. Corresponding behaviour of the force is evident in Figure \ref{fig:force} (also compare to Figure \ref{fig:aveN}).

\section{General case: outline of solution strategy}
\label{generalsection}
Suppose we return to the general form of the partition function in equation \eref{ztilde}. For the general case of an infinitesimally thin rod in the plane, the remaining integrals over $x$ are not as easy to diagonalize. Instead of such a rod we shall consider a flat slab of width $d$ lying along the $y$ axis in the $xy$ plane; see Figure \ref{fig:quadrant1}. This amounts to modifying the two-slit scenario of previous section by removing the ``outer barrier'' of each slit, thereby changing the $x$ integration domains to $(\frac d 2,\infty)$. The limit $d\rightarrow 0$ represents the original scenario of Edwards' rod in the plane.

Recall that the terms in the partition function consist of multiples of $T_c$ with even or odd geometric series of $T_s$, i.e., $T_c (1-T_s^2)^{-1}$ or $T_c T_s(1-T_s^2)^{-1}$. We approximate these contributions to the partition function in two steps:
\begin{itemize}
	\item address the sub-sequences of single-side contributions that originate from augmentations of the type \eref{rulea}, i.e., \\
	$ T_c(x+x_0) \underbrace{ T_s(x_0-x_1) \ldots  T_s(x_{m-1}-x_m)}_{m \;\mathrm{single-side\;terms}}  T_c(x_m+x'$), and then
	\item approximate the crossing terms $ T_c$.
\end{itemize}

\subsection{Approximation of $T_s$ sequences}
\label{tsapproxsect}
Let us begin by considering a sub-sequence of $m$ single-side contributions in the integral \eref{ztilde} between $x_0$ and $x_m$, say. For notational convenience we omit $k$ and $t$ dependence (which is the same in all terms), but recall that the $k$ integral still remains. Let us consider the case where we have \emph{not yet} performed the Laplace transformation w.r.t. $L$, and define
\beqa
{T_{\mathrm{eff.}}^{F}}(x_0,x_m;k;\{s_i\}) &=& \int_{d/2}^\infty dx_1\ldots \int_{d/2}^\infty dx_{m-1}\nl
&&\;\;\;\;  T_s^{F}(x_0-x_1;k;s_1)\ldots  T_s^{F}(x_{m-1}-x_m;k;s_m).
\label{teff1}
\eeqa
Here each $ T^{F}_s$ has been Fourier transformed in its $y$s but not yet Laplace transformed, i.e.,
\beq
T^{F}_s(x_i,x_{i+1};k;s_i) = \frac{\mathcal N}{s_i^2} e^{-3(x_i-x_{i+1})^2/2s_i - k^2 s_i/6},
\eeq
as is easily seen from \eref{tp}. This implies that we may write
\beq
\prod_{i=1}^{m-1} T^{F}_s(x_i,x_{i+1};k;s_i) = \left( \prod_{i=1}^{m-1} \frac{\mathcal N}{s_i^{3/2}}e^{- k^2 s_i/6}\right) \left( \prod_{j=1}^{m-1} \frac{1}{s_j^{1/2}} e^{-\frac{3(x_j-x_{j+1})^2}{2s_j}}\right).
\label{generalprod}
\eeq
We note that second product is simply one of Green functions for a one-dimensional polymer chain,
\beq
G_x(x_i,x_{i+1};s_i) = \frac{\mathcal N'}{s_i^{1/2}} e^{-3(x_i-x_{i+1})^2/2s_i}.
\eeq
The restriction imposed by the integration bound on each intermediate $x_i$ (where the polymer pierces the plane) implies that none of the piercings may enter the excluded region of the bar in the region $x \in (-\frac d 2,\frac d 2)$. This however does \emph{not} preclude any other part of the polymer arc to cross over this region, as illustrated in Figure \ref{fig:quadrant1}.
\begin{figure}[H]
	\centering
		\includegraphics[width=0.35\textwidth]{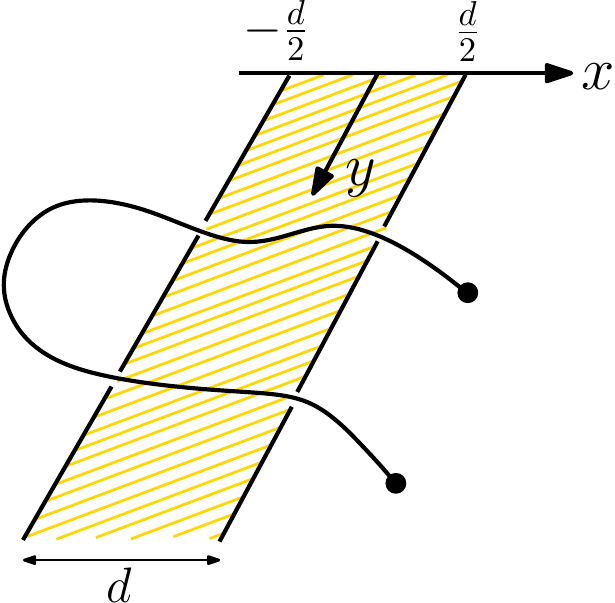}
	\caption{Piercings are excluded from a slab region on the $y$ axis, but the remainder of a polymer arc could still cross over this region.}
	\label{fig:quadrant1}
\end{figure}  
We proceed with our approximation by assuming that indeed \emph{none} of the segments in an arc may cross the excluded region, i.e., all $x$ co-ordinates of an arc (and not just those of its piercings) must lie in the region $(\frac d 2, \infty)$. Naturally  this approximation would exclude several possible configurations, and we are significantly under-estimating the true entropy of the polymer arc. As such this would provide a sensible upper bound for the free energy of the system. This case should be adaptable to the scenario of a polymer in a cavity from \cite{grosbergfrish2003} through a similar argument in a cylindrically symmetric setting.

Under this assumption the product $\prod_{i=1}^{m-1} G_x(x_i,x_{i+1};s_i)$ reduces to a single chain of length $\sum_{i=1}^{m-1} s_i$ that \emph{nowhere} crosses into the forbidden $x$ domain, i.e., the polymer arc is effectively restricted to a quadrant around the forbidden region. For such a restricted random walk we may once more use Chandrasekhar's argument of mirror images to construct an effective Green function,
\beqa
\prod_{i=1}^{m-1} G_x^{(x\geq d/2)}(x_i,x_{i+1};s_i) &=& G_x^{(x\geq d/2)}(x_0,x_{m};\sum_{i=1}^{m-1} s_i) \nl
&=& G_x^{(x\in \mathbf R)}((x_0-\frac d 2)-(x_{m} - \frac d 2);\sum_{i=1}^{m-1} s_i) \nl
&&\;\;- G_x^{(x\in \mathbf R)}((x_0-\frac d 2)+(x_{m} - \frac d 2);\sum_{i=1}^{m-1} s_i).
\label{Geffx} 
\eeqa
Here the superscripts on the Green functions refer to the respective $x$ integration domains. We have thus replaced the sequence of $G_x$s with an effective Green function by making use of the Markov property of random walks after having extended the $x$ integration to all space by making use of Chandrasekhar's argument. This is manifest in the last line above of equation \eref{Geffx} where the Green function with the reflected end co-ordinate has been subtracted. This means that all intermediate $x$ integrals have vanished, and only the integrals over the beginning and end $x$ co-ordinates of $G_x^{(x\geq d/2)}(x_0,x_{m};\sum_{i=1}^m s_i)$ remain. In this approximation the dependence on $m$ has completely disappeared out of one part of the expression in equation \eref{generalprod} yielding
\beq
\prod_{i=1}^{m-1} T_s^{F}(x_i-x_{i+1};k;s_i) \approx \left( \prod_{i=1}^{m-1} \frac{\mathcal N}{s_i^{3/2}}e^{- k^2 s_i/6}\right) G_x^{(x\geq d/2)}(x_0,x_{m};\sum_{i=1}^{m-1} s_i).
\eeq
What remains is to perform the integrals over the $s$ co-ordinates, with the relevant cut-off on the lower integration bound,
\beqa
\hat T_{\mathrm{eff.}}^{F}(x_0,x_m;t) &=&  \int_{-\infty}^\infty dx_0 \, dx_m\; \int_\epsilon^\infty \prod_{j=1}^{m-1} ds_j\nl
&&\;\;\;e^{-\sum_i s_i t} \prod_{i=1}^{m-1} \frac{\mathcal N}{s_i^{3/2}}e^{- k^2 s_i/6} \; G_x^{(x\geq d/2)}(x_0,x_{m};\sum_{i=1}^{m-1} s_i).
\eeqa
This effective quantity depends only on the first and last co-ordinates, $x$ and $x'$, and on the number $m$ of same-side steps in the sub-sequence. A summation over $m$ will lead to a geometric series $G^{\mathrm{odd/even}}_{\mathrm{eff.}}(x,x')$ that may now be included between $T_c$ terms in a chosen approximation scheme for the summation over valid sequences.

\subsection{Approximating the $T_c$ terms}

The second step is to deal with rod-crossing terms of the form $T_c(x,x')$. These terms are responsible for the localisation of the polymer around the rod at these points. We propose an approximation that decouples $x$ and $x'$, 
\beq
T_c(x,x') \approx f(x)f(x').
\eeq
This could be done in several ways, and would lead to a complete diagnoalisation of the integration. For a short polymer one may assume that the crossing terms will be localised close to the boundary of the slab, i.e., $\langle x \rangle \approx \frac d 2$. For other length scale regimes this average localisation could also be guessed as a function of the Laplace parameter $t$ or solved in some self-consistent manner.

A slightly more general approach would be to assume that the piercings of the crossing terms are localised by Gaussians around $x,x' = \frac d 2$. Naturally one would expect a $T_c$ term to show a corresponding decline if $x \in (\frac d 2,\infty)$ and $x'\in (\frac d 2,\infty)$ are moved further from the origin, since it is a Gaussian in $x+x'$. The Laplace parameter $t$ is related to the inverse of the arc-length of the crossing segment, and should be indicative of this localization. A sensible approximation for the $x$ dependent part would thus be to say that $T_c$ localizes $x$ and $x'$ independently (i.e., we decouple the function into two independent Gaussians),
\beq
T_c^{L}(x+x') \approx \mathcal{N}_q e^{-\frac{qt}{2}(x-\frac d 2)^2 -\frac{qt}{2}{(x'-\frac d 2)}^2},
\eeq
where the localization parameter $q$ is treated as a guess to the localisation length scale. Here $\mathcal{N}_q$ is the normalisation such that
\begin{equation}
\int_{0}^{\infty}dx\int_{0}^{\infty} dx'\,\,\mathcal{N}_q e^{-\frac{qt}{2}(x-\frac d 2)^2 -\frac{qt}{2}{(x'-\frac d 2)}^2} = 1.
\end{equation}

One could also determine the strength of the localisation using, for instance, a variational calculation. The result could now be combined with those of \sref{tsapproxsect} and \sref{sumsection} in some approximation for the partition function.

\section{Conclusions and closing remarks}

The problem of winding a polymer around an infinitely long obstacle was addressed by labelling configurations according to sequences of sub-arcs constrained to a half-space. By considering arcs that cross the obstacle or remain on the same side thereof, combinatoric rules were derived that allow winding number conserving augmentations of these sequences. These augmentations rely on topology conservation through type 2 Reidemeister moves. Properties of valid sequences were identified and an algorithm was presented for discriminating whether a given sequence is valid. 

Two possible approximations for the partition function were found by considering summations over valid configurations, one bounding the true partition function from above and the other from below. Given a particular choice of polymer variant (Gaussian, semiflexible etc.) a statistical weight may be attached to each sub-arc. Through a series of diagonalising transformations the lower bound approximation for the partition function was written as an integral of these statistical weights. 

For the case of a Gaussian chain a specific statistical weight was assigned, and the partition function for $w=1$ was approximated for the case of windings through two slits in a plane. For zero slit-width it was found that the basic undressed loop dominates the partition function when the polymer is short, as expected. Various expectation values were calculated and the free energy was plotted for different slit separations. The results make good physical sense for various average polymer length scales, lending credibility to our approximations.

The case of non-zero slit width was treated in a localising approximation. The force of the slits on the polymer shows an expected sign change as slit separation is varied for a fixed average polymer length. Correspondingly different length scale regimes arise that determine what fraction of the polymer is in crossing terms or in same-side terms.

Only the case $w=1$ was treated in detail. Since the partition function arises from summing integrals of products of functions, the partition function for higher winding numbers may in principle be constructed from higher order basic loops by our rules. This implies that some statistical quantities calculated from logarithms of $Z^{(w)}_{\mathrm{appx 1}}$ (e.g., $\langle L \rangle(t)$, $\langle N_s\rangle$, $\langle N_c\rangle$$\ldots$) scale with the winding number of the system. In our lower bound approximation scheme, at least, higher windings entail repeats of the integrand a corresponding number of times. The upper bound approximation scheme could also be useful in some regimes since the first few terms in the summation over configurations dominate the partition function for certain length scales.

Lastly, suggested approximations were sketched for a slab-like obstacle in the plane. It was observed that the latter case should be relatable to other confined geometries studied in existing literature. Our approach allows different and intuitive calculation strategies for partition functions of wound polymers.

\section{Acknowledgements}

C.M.R. would like to thank the Wilhelm Frank Bursary Trust for financial support and is grateful to Dr J.N. Kriel for several valuable discussions. K.K.M.-N. acknowledges funding from the National Research Foundation of South Africa.

\appendix

\section{Comments on enumeration and braid groups}
\label{braidappend}
We explain here briefly the relation of our enumeration procedure of section \ref{windingsection} to some properties of braid groups.
Since we wish to calculate partition functions and other averages for polymers of a given winding number we illustrate why braid group properties alone do not suffice to couple polymer degrees of freedom to a particular configuration.

A braid group $B_n$ describes braids of $n$ strings, and has $n-1$ group generators $\{\sigma_1,\sigma_2,\ldots,\sigma_{n-1}\}$. The generators obey the relations
\beqa
\sigma_i&\sigma_{i+1}\sigma_i = \sigma_{i+1}\sigma_i\sigma_{i+1}, \qquad i = 1,2,\ldots,n-1, \nl
\sigma_{i}&\sigma_j =\sigma_{j}\sigma_i, \qquad\qquad\qquad |i-j|\geq 2, \nl
\sigma_{i}&\sigma_i^{-1} = \sigma_i^{-1} \sigma_{i} = e.
\label{braidgroup}
\eeqa
Here $e$ is the identity element. One may construct words from this set of generators and their inverses; these words correspond to braids of the $n$ strands. (See, for instance, \cite{nechaev1996,nechaev1999,kauffman,manturov} for further details.) For any given word the application of the braid group relations (\ref{braidgroup}) may be used to obtain a minimal (irreducible) form of word. 

The scenario of winding a strand around a rod could, of course, be viewed in terms of the braid group for two strands, $B_2$. This group is ``trivial'' in that it only has one generator $\sigma$, rendering the first two properties of (\ref{braidgroup}) irrrelevant. 
\begin{figure}[h]
	\centering
		\includegraphics[width=0.95\textwidth]{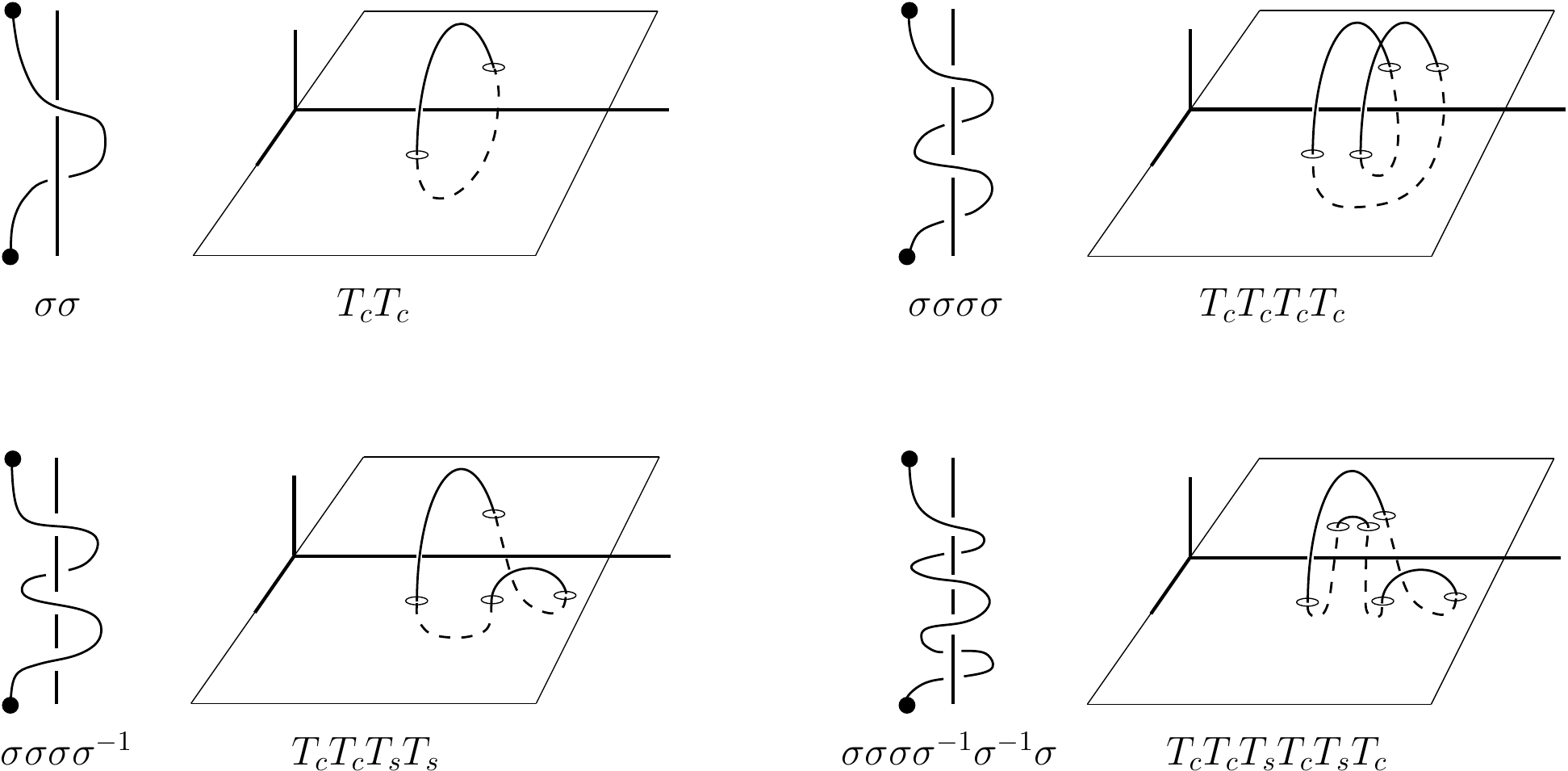}
		\caption{Braids of two strands in analogy to winding scenarios from section \ref{windingsection}.}
	\label{fig:braid1}
\end{figure}
In Figure \ref{fig:braid1} we illustrate the $B_2$ words analogous to some sequences of $T$s from section \ref{windingsection}. The braids shown there are projections of the winding scenarios along the plane in which the rod lies. Clearly the group properties thus encode similar conditions on reducibility as set out in section \ref{windingsection}. Since we consider words of even length, the winding number of a particular $B_2$ braid may be obtained by halving the number of $\sigma$ generators remaining in the corresponding word that has been fully reduced. For any word of even length, the winding number is clearly related to the difference in number of $\sigma$s and $\sigma^{-1}$s,
\beq
w = \frac {\#(\sigma) - \#(\sigma^{-1})} 2.
\label{braidwind}
\eeq
(Compare to the power of the Alexander invariant in equation (2.58) of \cite{nechaev1996}.)

However, our aim is to couple polymer degrees of freedom (statistical weights) to these words. These weights are simply the probability distributions of connected arcs constrained to half-space, \emph{viz.} the $T$s. It is critical to identify the correct integration bounds on the beginning and end co-ordinates in these weights (see sections \ref{tdetails} and \ref{diagsect}) when stringing together these $T$s for a given braid / winding scenario. Indeed, the group relation $\sigma \sigma^{-1}= \sigma^{-1} \sigma$ is problematic in that it alone does not, for instance, allow us distinguish between the sequences $T_s T_s T_s$ and $T_c T_s T_c$. Clearly, through (\ref{ruleanew}) and (\ref{rulebnew}) both of these sequences reduce to $T_s$. In our partition function, however, where the $T_c$s and $T_s$s do not have the same integration bounds, these two configurations would represent different statistical contributions.

Naturally it is possible to enumerate all unique (non-primitive) words in the ``braids only'' scenario, where the winding number constraint $w=1$ may analogously be viewed as enumerating all words in $B_2$ that reduce to $\sigma \sigma$. One could do this by enumerating all possible words of even length, and then enforcing $w=1$ through (\ref{braidwind}) and an appropriate Kronecker delta. However, the translation to arcs with position dependence must be made explicit to avoid the above-mentioned problem. The group relations alone (without position dependence) thus do not allow coupling to the polymer degrees of freedom. 
Our scheme has thus been developed to avoid the necessity of including the $\#(\sigma) - \#(\sigma^{-1})$ constraint in terms of an additional integral for the Kronecker delta by enumerating the configurations (with position dependence) explicitly.  The sequence of $T$s does matter since it indicates the order of coupling of positional degrees of freedom.

\section{Redundancy of rule (i'b)}
\label{summationappendix}

To see why rule (i'b), for example, is not necessary when using rule (i'a), we consider how an initial knot configuration, given by a pure winding number such as
\begin{equation}
Z^{(w)}_{\mathrm{basic}} = \left( T_c T_c \right)^w
\end{equation}
is modified by the application of these two moves.  Inserting rule (i') after the second $T_c$ is equivalent to inserting $T_c T_s T_c T_s$ after the first $T_c$.  Further equivalence can also be avoided by noting that the composite rules produce products of terms of the form 
\begin{displaymath}
T_c \rightarrow \left(
T_c T_s T_c T_s + T_s T_c T_s T_c
\right)^m.
\end{displaymath}
This can simplified by noting that the following cross product terms arise, but can be derived from from (ii') with the odd/even $T_s$ convention
\begin{displaymath}
T_c T_s T_c T_s \times T_s T_c T_s T_c \leftarrow 
T_c T_s \underbrace{T_c \underbrace{T_s.}_{(ii')}}_{\mathrm{then \, even \, }T_s\mathrm{ \,after\, 1st \,}T_c}
\end{displaymath}

\section{Various transformations in section \ref{diagsect}}
\label{diagappend}

The general form of \eref{zgeneral} may be simplified by performing a Laplace transformation. Suppressing all $x$ and $y$ dependences, we condense \eref{zgeneral} to
\beqa
Z^{(w=1)}_\chi(L) =& \int_0^\infty dX \int_{-\infty}^\infty dY \int_\epsilon^\infty ds_1\ldots \int_\epsilon^\infty ds_N \;\;\delta(x_0-x_N)\delta(y_0-y_N)\nl
&\qquad T_{p_1}(s_1)\ldots T_{p_N}(s_N) \;\delta(\sum_{k=1}^N s_k-L).
\eeqa
Performing the Laplace transformation (indicated by the tilde below) yields
\beqa
\tilde Z^{(w=1)}_\chi (t) &=& \int_0^\infty dL \;\;Z^{(w=1)}_\chi(L) \;e^{-Lt} \nl
&=&\int_0^\infty dX \;\delta(x_0-x_N)\int_{-\infty}^\infty dY\;\delta(y_0-y_N)\; T^{L}_{p_1}(t) \ldots  T^{L}_{p_N}(t),
\label{laplace}
\eeqa
i.e., by performing all integrals over $s$ we obtain the product of the Laplace transformations of the $N$ individual $T$s. The superscript $L$s indicate the Laplace transform of \eref{tp}. It is, however, understood that all $s$ integrals are performed from the non-zero lower bound $\epsilon$. The object \eref{laplace} may further be simplified in terms of the $y$ integrals. To this end we perform the linear and invertible co-ordinate transformation $\Delta y_i \equiv y_i-y_{i+1}$, $i=0,1,\ldots N-1$ and $R \equiv \sum_{i=0}^N y_k$. The Jacobian is $|J| = N+1$. Now suppressing the $x$ and $t$ dependences we may write \eref{zgeneral} as
\beqa
\tilde Z^{(w=1)}_\chi (t) &=& \int_0^\infty dX \;\delta(x_0-x_N) \int_{-\infty}^\infty dy_0\ldots \int_{-\infty}^\infty dy_N \nl
&&\qquad T^{L}_{p_1}(y_0-y_1)\ldots  T^{L}_{p_N}(y_{N-1}-y_N) \;\delta(y_0-y_N) \nl
&=& |J|\int_0^\infty dX \;\delta(x_0-x_N)\int_{-\infty}^\infty d\Delta y_1\ldots \int_{-\infty}^\infty d\Delta y_N \int_{-\infty}^\infty dR \nl
&&\qquad  T_{p_1}^{L}(\Delta y_1)\ldots   T_{p_N}^{L}(\Delta y_N) \;\delta(\sum_{k=1}^N \Delta y_k) \nl
&=& \mathcal N \int_0^\infty dX\;\delta(x_0-x_N) \int_{-\infty}^\infty dk\;\;  {T}_{p_1}^{L,F}(k)\ldots  T_{p_N}^{L,F} (k).
\label{fourier}
\eeqa
where $\mathcal N= |J| \int_{-\infty}^{\infty}dR$ is a pre-factor related to the length of the rod. This pre-factor is, in principle, divergent, but its contribution to the free energy is additive and thus irrelevant. The superscripts in the final line above indicate Fourier transformations in $y$ of the Laplace transformations of the $T$s from \eref{tp}. We used the Fourier representation of the delta function,
\beq
\delta(\sum_{k=1}^N \Delta y_k) = \int_{-\infty}^\infty dk\;\; e^{ik\,(\sum_{k=1}^N \Delta y_k)},
\label{fourierdelta}
\eeq
to do all the $\Delta y$ integrals. Recalling  that in \eref{fourier} we suppressed dependence on the Laplace parameter $t$ and on the various $x$ co-ordinates, we write the Laplace transformation of $Z^{(w=1)}_\chi$ explicitly as
\beq
\tilde Z^{(w=1)}_\chi (t) = \int_0^\infty dX \;\delta(x_0-x_N)\int_{-\infty}^\infty dk\;\; \prod_{i=1}^N  T^{L,F}_{p_i}(x_{i-1},x_i;k;t),
\eeq 
as in \eref{ztilde}.

\newpage
\section*{References}
\bibliographystyle{unsrt}
\bibliography{references}

\end{document}